\documentclass{article}
\usepackage{graphicx}
\usepackage{natbib}
\begin{document}

\begin{center}
\Large
{\bf A Heavy Baryonic Galactic Disc}
\large

{\bf J. I. Davies}

{School of Physics and Astronomy, Cardiff University,\\ The Parade, Cardiff CF24 3AB, UK, \\(Jonathan.Davies@astro.cf.ac.uk) \\
April, 2012.}
\end{center}
\normalsize

\begin{center}
{\bf Abstract}\\
\end{center}
We investigate the possibility that the observed rotation of galaxies can be accounted for by invoking a massive baryonic disc with no need for non-baryonic dark matter or a massive halo. There are five primary reasons for suggesting this possibility. Firstly, that there are well known disc surface mass density distributions that naturally produce the observed rotation curves of galaxies. Secondly, that there are a number of rotation curve `puzzles' that cannot be explained by a massive dark matter halo i.e. the success of maximum disc fitting, HI gas scaling to the observed rotation, the disc/halo conspiracy and the interpretation of the Tully-Fisher relation. Thirdly, recent 21cm observations show an almost constant HI surface density and a distinct `cut-off' or edge to galactic discs. We explain this constant surface density in terms of either an optical depth effect or the onset of molecular gas formation and hence the possibility of considerably more gas existing in galaxies than has previously been thought. We suggest that the HI cut-off does indeed mark the edge of the galactic disc. Fourthly, there have recently been an increasing number of observations, most importantly $\gamma$-ray observations, that imply that $X_{CO}$ may be ten times higher in the outer Galaxy. This `dark' gas may provide adequate mass to account for galaxy rotation. Finally, we show that the additional baryonic mass required to account for the rotation of galaxies is just that required to reconcile observed baryons with those predicted by big bang nucleosynthesis without having to invoke a massive warm inter-galactic medium. We reconsider Mestel's ideas about the collapse of isothermal and constant density spherical clouds and show that these can be simply used to successfully model the rotation of galaxies. Mestel discs can also be used to straight forwardly explain the scaling laws of galaxies, particularly the observed relation between rotation velocity and radius and the oft used Tully-Fisher relation. Thus the observed gross properties of disc galaxies can be explained by the monolithic collapse of baryonic gas into a rotationally supported disc. We discuss observations of the baryonic content of galactic discs and where sufficient `hidden' baryons might be found to account for the rotation. 

\section{Introduction}
Over the last 30 years dark matter has played an increasingly important role in our interpretation of Galactic, extra-galactic and cosmological observations. Circumstantial evidence for dark matter has gradually been accumulating from a wide variety of disparate observations; from locally observed large scale structure (Doroshkevich et al., 2004) to the distant signals imprinted in the cosmic microwave background (Spergel et al., 2007). In this paper we want to take what some might describe as a `blinkered' but, we hope, insightful look at one specific aspect of the dark matter problem. We discuss the requirement for a dark matter halo around galaxies and specifically whether observations of galaxies could be explained more satisfactorily by a massive baryonic disc without resorting to non-baryonic dark matter at all. We do not address other evidence for dark matter. Our intention is to tackle a dilemma. Experiments have failed to detect dark matter particles, particle physics has failed to provide convincing  dark particle candidates and modifications to gravity do not sit comfortably within well tested general relativity. Perhaps one aspect of the dark matter problem, galactic rotation, can be explained by tangible well observed and familiar baryons without resulting to some new unknown form of matter.
 
The almost constant velocity of rotation of galaxies at large distances from their centres (flat rotation curves) is most often quoted as the primary evidence for dark matter. In fact flat rotation curves in themselves provide no evidence at all because there is a disc surface mass distribution (see below) that naturally produces a flat rotation curve. What is important is not that the curve is flat, but the value of the asymptotic velocity and the large distances from the galactic centre that this velocity is measured. For a typical spiral galaxy, the asymptotic velocity at large distance from the centre is too great to be accounted for by the sum of the known baryonic components (stars, gas, dust) - hence the inference of dark matter.

Before dark matter came into vogue astronomers were content to try and model galactic rotation using only the observed baryonic material. A seminal paper on this subject was written by Mestel (1963) (see also Hunter et al., 1984). The paper is quite extraordinary and brings into question the normally accepted and often quoted history of galaxy rotation, which firmly places the discovery of flat rotation curves in the 1970s, to early 1980s (Rogstad and Shostak, 1972, Roberts and Rots, 1973, Rubin et al., 1978)
\footnote{Though note Babcock's comment with regard to M31 that '...a very great proportion of the mass of the nebula must lie in the outer regions.', Babcock (1937).}. Mestel was trying to derive the surface mass density of two types of galactic disc that were distinguished by their observed rotation curves. The first had constant angular velocity (solid body rotation) and the second constant linear velocity. These two forms are just what are observed - many but not all galaxies have flat rotation curves (Rubin et al., 1978). 

Mestel showed that these two rotation laws can arise from two quite simple disc surface mass density distributions.
Constant angular velocity:
\begin{equation}
\Sigma(r)=\Sigma_{0}\left[1-\left(\frac{r}{R_{0}}\right)^{2}\right]^{1/2}
\end{equation}
and constant linear velocity:
\begin{equation}
\Sigma(r)=\frac{\Sigma_{0}R_{0}}{r}
\end{equation}
where $\Sigma$ is the mass surface density, $r$ the radial distance from the centre of the disc and $R_{0}$ the size of the disc ($r \le R_{0}$). (2) above is often referred to in the literature as Mestel's disc and is not exact (see below) because of the problem with the discontinuity at $R_{0}$. (1) above is what you would expect in the limit as a uniform density spheroid shrinks one of its axes to zero. This is important because Mestel's intention was not only to derive the surface mass distributions, but to also see if they were the viable result of the gravitational collapse of a cloud of gas - the then preferred mechanism for galaxy formation. He demonstrated that both are acceptable solutions to the collapse problem provided the angular momentum distribution function ($M(h)$) is preserved during the collapse i.e. that viscous transport is negligible - if so then (1) is the result of a collapsing uniform density cloud and (2) arises from a cloud in which the density `decreases slightly' with increasing radius.

$M(h)$ is the function that specifies the amount of mass with angular momentum less than $h$. This function provides a test of Mestel's idea and ultimately an insight into the galaxy formation process. Crampin and Hoyle (1964) were the first to test Mestel's hypothesis that $M(h)$ was the same for disc galaxies as it was for a uniformly rotating spheroid (the pre-galactic gas cloud). Using eight galaxies they confirmed Mestel's result. Essentially they measured the rotation and surface density as a function of radius to derive $M(h)$. Freeman (1970) extended this work to a larger sample and came to the same conclusion as Crampin and Hoyle. That there was agreement between observation and Mestel's theory was in some ways fortuitous and in some ways insightful. It was fortuitous because at that time essentially only the inner solid-body part of the rotation curve was being measured. Also Freeman showed that $M(h)$ for an exponential surface density distribution was very similar to the $M(h)$ of Mestel's disc, thus importantly demonstrating that $M(h)$ alone is not sufficent to determine the surface mass distribution of a self-gravitating disc. It was insightful because it provided something we no longer have - consistency between the observed distribution of matter (baryonic), how that matter rotates and a theory of how galaxies form. 
Today observations of disc galaxies have moved on, but maybe not forward. With a relatively high velocity flat rotation curve and a relatively low mass exponentially distributed baryonic disc $M(h)$, at first sight, no longer provides a link to a plausible pre-galactic gas cloud.

Our current theoretical ideas about galaxy formation are now quite different to those of Mestel. The hierarchical theory (White and Rees, 1978) places much more emphasis on galaxy merging as the process of large galaxy formation, rather than a single monolithic collapse. However, although there is clear evidence that galaxy merging does occur there is no clear evidence, for a galaxy like the Milky Way, that it is the major physical process at the root of its formation. Eggen, Lynden-Bell and Sandage (1962), using the metalicities and velocities of stars, were the first to present us with good evidence that the Galaxy is the result of the gravitational collapse of a proto-galaxy - a slowly rotating cloud reduced to a thin disc within about $10^{9}$ years. A picture of early star formation in a spheroid (Population II) and subsequent star formation in a metal enriched thin disc (Population I) is still a viable explanation for at least some part of what we call the Milky Way galaxy. Based on the thickness of the Galactic disc, Toth and Ostriker (1992) have argued that less than 4\% of the mass within the solar circle can have come from mergers and that this is inconsistent with the hierarchical galaxy formation models. This problem with the thinness of galactic discs has continued to provide a challenge to those who want to make galaxy mergers the major galaxy assembly mechanism (Navarro et al., 1994, Benson et al., 2004). 

If, as we believe, gravitational collapse still provides a viable explanation of the way a galaxy initially forms then might a Mestel disc be a viable explanation of the way it rotates, without the need for dark matter or alterations in the law of gravity? In this paper we want to reconsider Mestel's ideas. The posibility of a massive baryonic Mestel disc is the first motivation for this paper.

Our second motivation arises from some well known and long standing puzzles regarding the rotation of galaxies.
A rotation curve is normally modelled as the sum of the velocities of the presumed component parts of a galaxy:
\large
\begin{center}
$v^{2}_{C}=v^{2}_{Stars}+v^{2}_{Gas}+v^{2}_{DM}$
\end{center}
\normalsize
Each component's contribution to the circular velocity $v_{C}$ is derived from the surface density of stars and gas and then the contribution of the dark matter is added so that the sum corresponds to what is observed. The dark matter is normally assumed to be distributed in an approximately spherical halo while the stars and gas are in a disc. We, along with others before us (Pfenniger, Combes and Martinet, 1994 and references therein), identify four puzzles.
\begin{enumerate}
\item {\bf Gas disc scaling} - It was Bosma (1978, 1981) who first noted that to account for rotation the required total surface density of matter in the spiral discs he studied was, in the outer regions, approximately proportional to the surface density of atomic hydrogen. In fact `wiggles' in the rotation curves of some galaxies are directly reproduced by the inferred rotation predicted by the atomic gas alone (Fig. 1). There seems to be an intimate relation between the atomic gas and the dark matter.
\begin{figure}
\begin{center}
\includegraphics[scale=0.45]{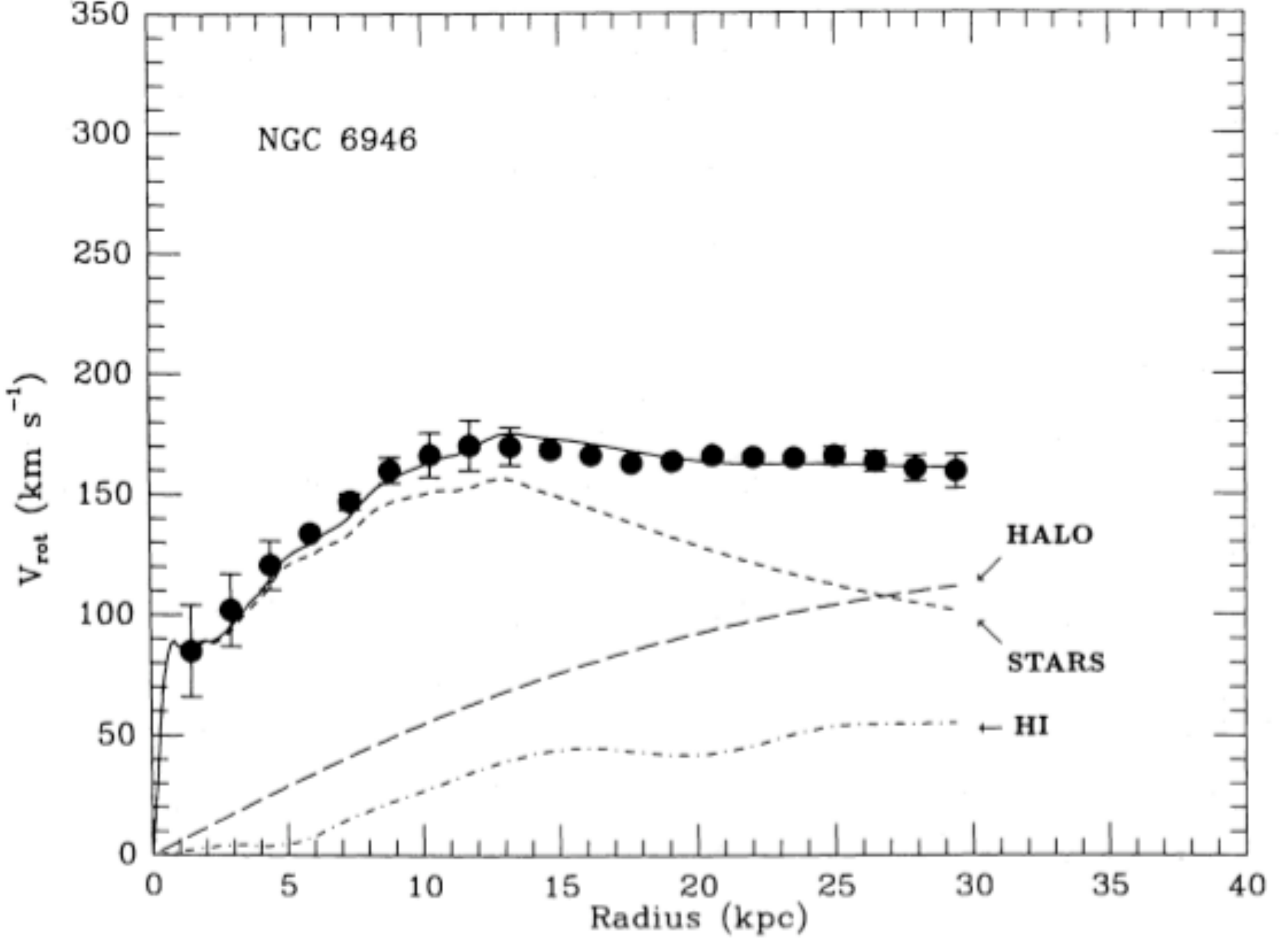}
\caption{The lower dot-dashed line is the rotation predicted for the gas alone. The upper black data points are the measured rotation curve. Note how the shape and even the `wiggle' in the gas are reproduced in the observed rotation curve. The data shown are for NGC6946 - taken from Carignan et al. (2008).}
\end{center}
\end{figure} 
\item {\bf Maximum disc fitting} - a problem with predicting the rotation due to the stars alone is the value of the mass-to-light ratio to use. Different age stellar populations and an imprecise knowledge of the initial stellar mass function can lead to quite different mass-to-light ratios. A solution has been to use the maximum disc model where the mass-to-light ratio of the stars is scaled so that the predicted rotation due to the stars fits the observed rotation curve in the inner regions. The deficit in rotation at larger radii is made up by the gas and dark matter (Palunas and Williams, 2000). The required maximum disc scaling is often consistent with expected stellar mass-to-light ratios and also with the stability of the stellar disc and the formation of spiral arms (Fuchs 2002). Freeman (1992) showed that maximum discs provide an excellent fit to rotation curve data - there does not appear to be any need for dark matter in the central regions of a galaxy. 
\item {\bf The disc halo conspiracy} - Given that over the inner region of a galaxy the stars and gas in a disc contribute significantly to the rotation while it is the spherical dark halo that does this in the outer regions, it is very strange that they `conspire' to produce a flat rotation curve (van Albada and Sancisi, 1986). There is no known physical reason for why this might happen.
\item {\bf The Tully-Fisher relation} - The Tully-Fisher relation is one of the best-known scaling relations for galaxies (Tully and Fisher, 1977). It was originally proposed as a means of obtaining distances to galaxies for comparison to their Hubble velocity distance, so that peculiar velocities could be derived. It has subsequently been used as a scaling relation in numerical simulations to infer luminosity from measured dark halo rotation. The Tully-Fisher relation is between the observed velocity width derived from the 21cm line profile ($\Delta v$) and the luminosity ($L$). If galaxies are dark matter dominated then the line width is related to the mass ($M$) and radial extent ($R$) of the dark matter while the luminosity is related to the baryonic material. The dark matter and luminous material again seem to have an intimate knowledge of each other. One interpretation of the Tully-Fisher relation is obtained by considering dynamical equilibrium:
\large
\begin{center}
$\Delta v^{2} \approx \frac{GM}{R}$ or $\Delta v \propto \left(\frac{M}{R}\right)^{1/2}$ \\
\end{center}
\normalsize
Now if $M=L\left(\frac{M}{L}\right)_{\odot}=LQ$ and $\Sigma=\frac{L}{R^{2}}$ then
\large
\begin{center}
$\Delta v \propto L^{1/4}Q^{1/2}\Sigma^{1/4}$
\end{center}
\normalsize

In this case $\Sigma$ is now the surface brightness and $Q$ the mass-to-light ratio. From the above the slope of the Tully-Fisher relation is predicted to be 1/4 as long as $Q$ and $\Sigma$ are constant. The slope is observed to be, as predicted, approximately 1/4, but Zwaan et al. (1995) have shown that galaxies with a wide range of surface brightnesses obey the same Tully-Fisher relation - there seems to be some conspiracy such that as surface brightness decreases the mass-to-light ratio increases in exact proportion. There is also no known reason why this should be so.
\end{enumerate}

It is difficult to see how these four enigmas can be accommodated within our current theories of galaxies, galaxy formation and dark matter. 

\begin{figure}
\begin{center}
\includegraphics[scale=0.65]{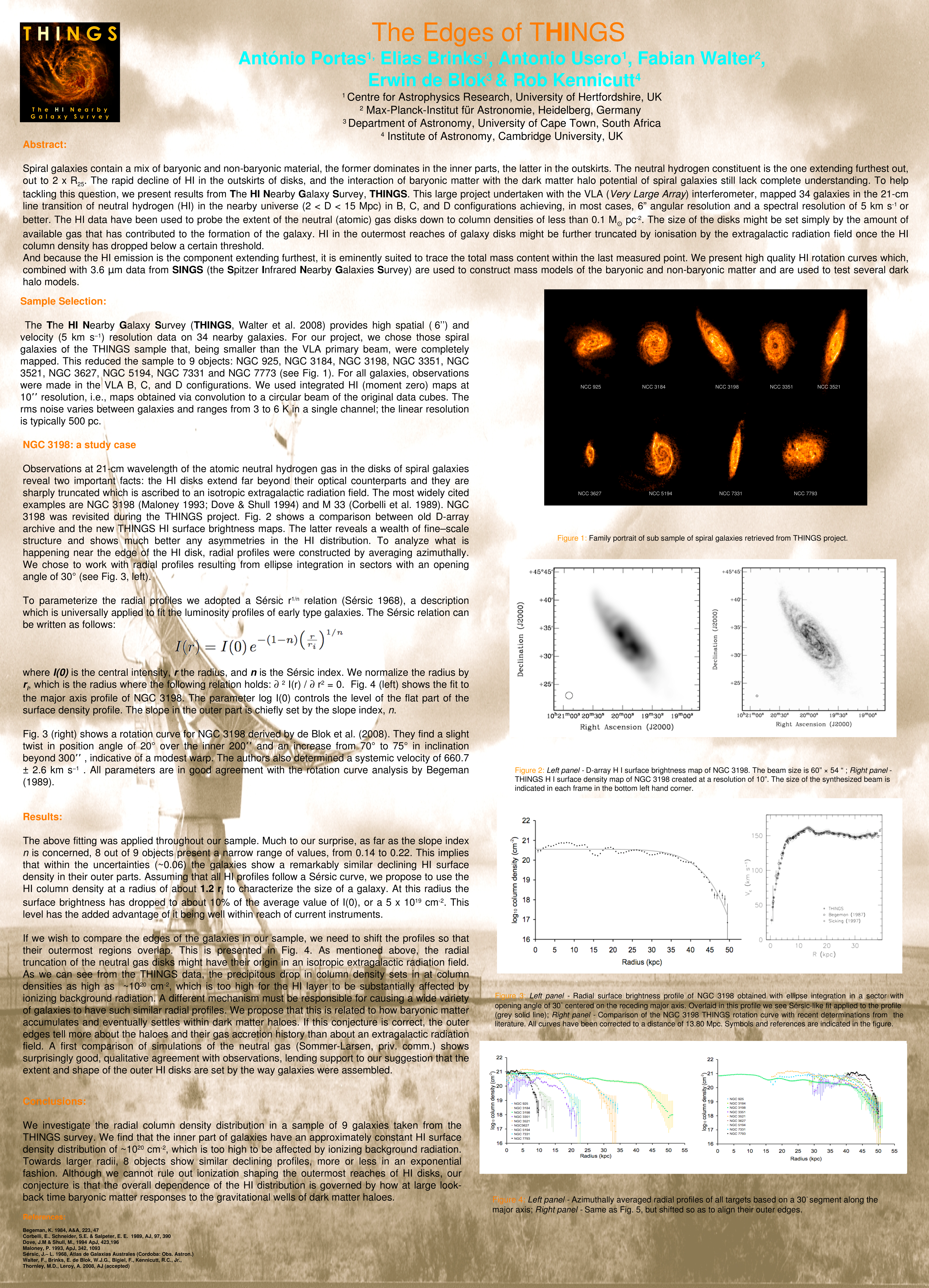}
\caption{The HI surface density of THINGS galaxies scaled to the same maximum radius - taken from Portas et al., (2008).}
\end{center}
\end{figure}

Our third motivation for this paper comes from recent observations of atomic gas in galaxies. The THINGS team have been carrying out some of the most sensitive interferometric observations of galaxies using the 21cm line of atomic hydrogen (Walter et al., 2008). These detailed observations (Fig. 2) illustrate some remarkable similarities for galaxies, which cover a large range in optical properties. The first intriguing thing is that they all have very similar HI column densities of order $10^{20-21}$ cm$^{-2}$, which remains reasonably constant with increasing radius over the entire disc (Bigiel et al. 2008, see also Minchin et al. 2004).
The THINGS team's explanation of this is that at column densities above $\sim 10^{21}$ cm$^{-2}$ the atomic hydrogen is converted into molecular hydrogen (H$_{2}$). Interestingly, a column density of $\approx10^{21}$ cm$^{-2}$ is also just about the point at which atomic hydrogen becomes optically thick (see below) and so as well as molecules, there is also the possibility of more atomic gas hiding in the disc. The THINGS surface density profiles also show that there is an edge to the HI - a striking and rapid decline in the HI column density (Fig. 2). The THINGS team interpret this as the onset of ionisation of the HI by inter-galactic ultra-violet radiation. We will show below that the column densities of the discs of these galaxies are too high for such a sudden decline to be caused by ionisation. We alternately suggest that this drop in HI column density really does mark a physical edge of the galactic disc (Bland-Hawthorn et al. 1997).

In summary, Mestel's ideas about gravitational collapse and the formation of a disc that has a flat rotation curve seem relatively simple, elegant and applicable to a galaxy like the Milky Way. There are a number of disturbing puzzles about the way galaxies rotate that are difficult to explain if galaxy masses are dominated by a dark matter halo rather than just material in a disc. Finally, does constant HI surface density provide a clue with regard to the distribution of baryons within the disc and does the sharp cut-off in the HI mark the edge of the disc? In this paper we explore the hypothesis that galaxies are massive gas discs, so massive that their rotation can be accounted for by baryonic material alone.  

Note that many of the issues highlighted in this paper have previously been discussed in Pfenniger, Combes and Martinet, 1994
(see also Pfenniger and Combes, 1994) and more recently by Hessman and Fiebart (2011). We hopefully provide further insight into essentially the same observational conundrums.

\section {Testing Mestel's ideas about gravitational collapse}
Mestel's calculations provide a simple, but eloquent analytical solution to the disc formation problem.
We will now consider Mestel's conclusions in a little more detail and solve the equations numerically, that he looked to solve analytically. First we will consider the assumptions that go into his model. 
\begin{enumerate}
\item A primeval gaseous sphere of 'roughly' uniform density ($\rho$) rotating with angular velocity $\Omega$ contracts under its own gravity with each element conserving its angular momentum. So, the collapse occurs purely along the $z$ direction i.e. parallel to the rotation axis. 
\item The turbulent and thermal motion of the gas make small contributions to the momentum balance and so the centrifugal force acting on the gas is a measure of the radial gravitational field near to the galactic plane. 
\item The change in gravitational energy is dissipated in shocks and radiated away. 
\item No sub-condensations are formed and there is no further supply of kinetic energy by for example hot young stars. 
\item The collapse continues until a thin disc is formed and the infinitely thin disc approximation can be used. This requires that: 
\large
\begin{center}
$\frac{GM}{R_{0}}>>c^{2}$ 
\end{center}
\normalsize
which is true for a typical galaxy as long as the gas temperature is below about $10^{5}$ K ($M\approx10^{11}$ $M_{\odot}$, $R_{0}\approx30$ kpc, $c$ is the velocity dispersion of the gas).
\item If the surface density changes with radius then so does the volume density and there will be a pressure gradient, but the force that this exerts is much smaller than that due to rotation and so it is ignored.
\item The gravitational field perpendicular to the rotation axis changes so that centrifugal balance is maintained. 
\end{enumerate}
Mestel's intention was to derive  equilibrium distributions of mass and angular momentum of the disc given the conditions in the primeval sphere. The equilibrium rotation field is then: \\
\large
\begin{center}
\begin{equation}
\Omega^{2}=-\frac{1}{r}\frac{d\phi}{dr}
\end{equation}
\end{center}
\normalsize

The potential at $r$ of a thin disc with size $R_{0}$ and surface density $\Sigma$ is given by:
\large
\begin{center}
\begin{equation}
\phi(r)=G \int^{2\pi}_{0} \int^{R_{0}}_{0} \frac{\Sigma(r_{p})r_{p}}{(r^{2}+r_{p}^{2}-2rr_{p}\cos{\theta})^{1/2}} d\theta dr_{p}
\end{equation}
\end{center}
\normalsize

In each of the cases below we will assume a mass of $M=3\times10^{11}$ M$_{\odot}$ and a radius of $R_{0}=30$ kpc - a galaxy similar to the Milky Way. Let us first consider the collapse of the uniform density sphere. In this case the surface density is trivially:
\large
\begin{center}
$\Sigma(r)=\Sigma_{0}\left[ 1-\left(\frac{r}{R_{0}}\right)^{2}\right]^{1/2}$
\end{center}
\normalsize
After considering the gravitational force exerted by an ellipsoidal mass distribution at a point on its equator as it becomes more flattened Mestel concluded that the resultant equilibrium disc would rotate with a constant angular velocity of:
\large
\begin{center}
$\Omega_{0}^{2}=\frac{3\pi}{4}\frac{GM}{R_{0}^{3}}$
\end{center}
\normalsize
In Fig. 3 we compare Mestel's prediction (blue dashed line) with that obtained by numerically integrating eq. 2 to get the potential and then using eq. 3 to get the linear velocity $\Omega r$ (blue solid line). The discrepancy between the two is because Mestel's calculation is derived from the limit of a spheroid that shrinks one of its axis to zero and as such is not quite the same as doing the integration (numerically) over a finite disc. The conclusion though is essentially the same i.e. the collapse of a uniform density spherical cloud leads to a disc that rotates with approximately constant angular velocity. The surface density of the resultant disc is shown in Fig. 4, of particular interest (see below) is the almost constant surface density within the inner regions.

\begin{figure}
\centering
\includegraphics[scale=0.55]{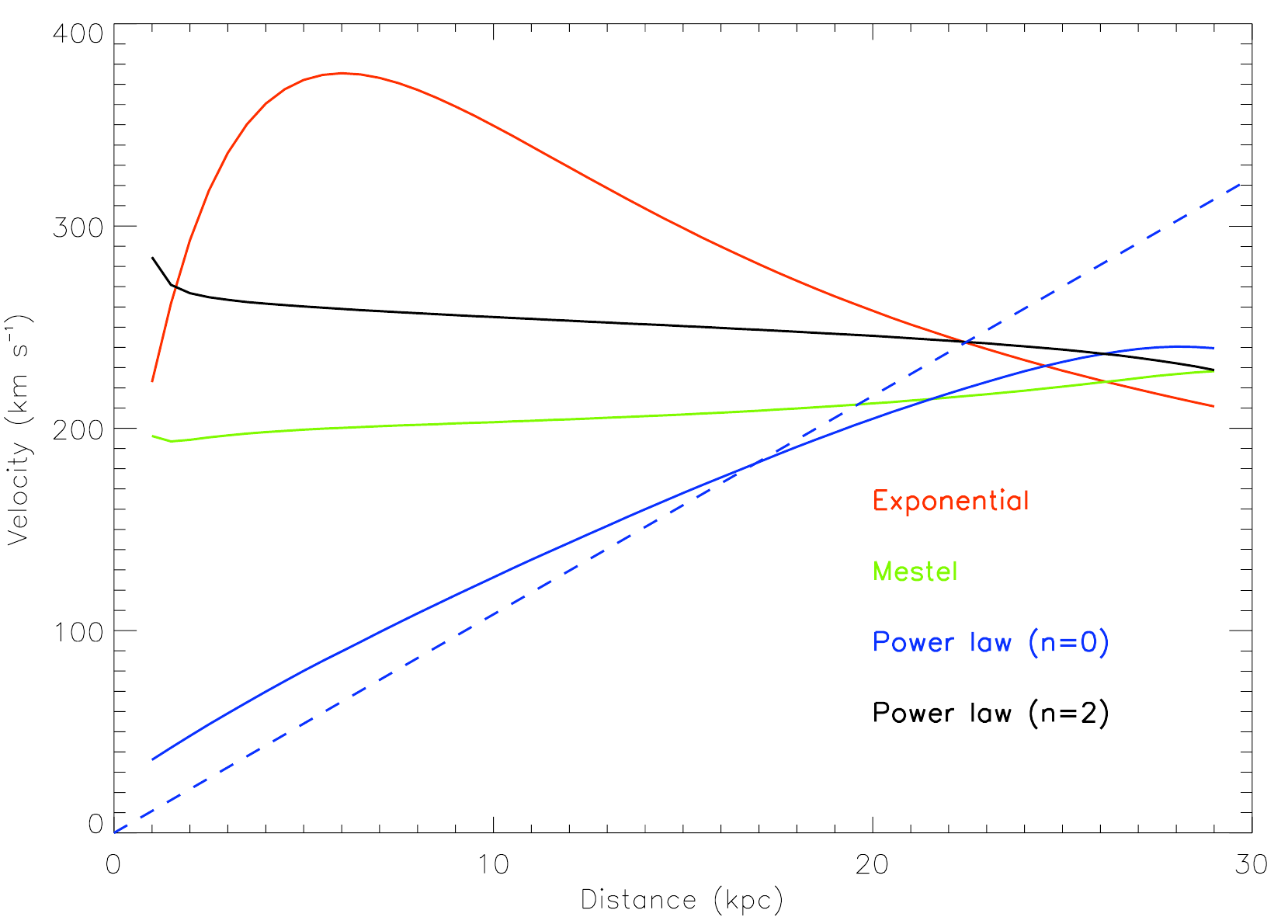}
\caption{The equilibrium rotation curves, calculated numerically, for the collapse of a uniform (blue) and power law density ($n=2$) (black) spherical cloud. The blue dashed line is Mestel's (analytic) prediction of the constant angular velocity resulting from the collapse of the uniform density sphere. The 'Mestel disc' and exponential rotation curves are shown in green and red respectively. In each case the disc has a mass of $M=3\times10^{11}$ M$_{\odot}$ and a radius of $R_{0}=30$ kpc, the exponential distribution has a scale length of 3 kpc.}
\end{figure}

The other interesting case is that of `Mestel's disc' and what density spheroid gives rise to it? In his paper Mestel comments that a $1/r$ surface density distribution results from the collapse of a spheroid that has a `slowly varying density', he does not quantify this. We have experimented with initially spherical power law  mass density distributions to get the best chi squared fit to a $1/r$ surface density law, there being no exact fit using a power law. To two decimal places we find that $n=1.76$ gives the best fit to a $1/r$ surface density distribution given the spheroidal mass density distribution $\rho(w)=\rho_{0}(w/R_{0})^{n}$, where $w$ is the radial co-ordinate of the spheroid. This is very close to the $n=2$ case which is the well known density distribution for the singular isothermal sphere i.e the collapse of a singular isothermal sphere gives rise to a surface density distribution that is very close to a Mestel disc ($\Sigma(r)=\Sigma_{0}(r/R_{0})^{n_{s}}$ with $n_{s}\approx-1$). In Fig. 4 we show the resultant surface density distribution ($n=2$) along with that of the exact ($1/r$) Mestel disc (black and green lines), they are very similar. 

\begin{figure}
\centering
\includegraphics[scale=0.55]{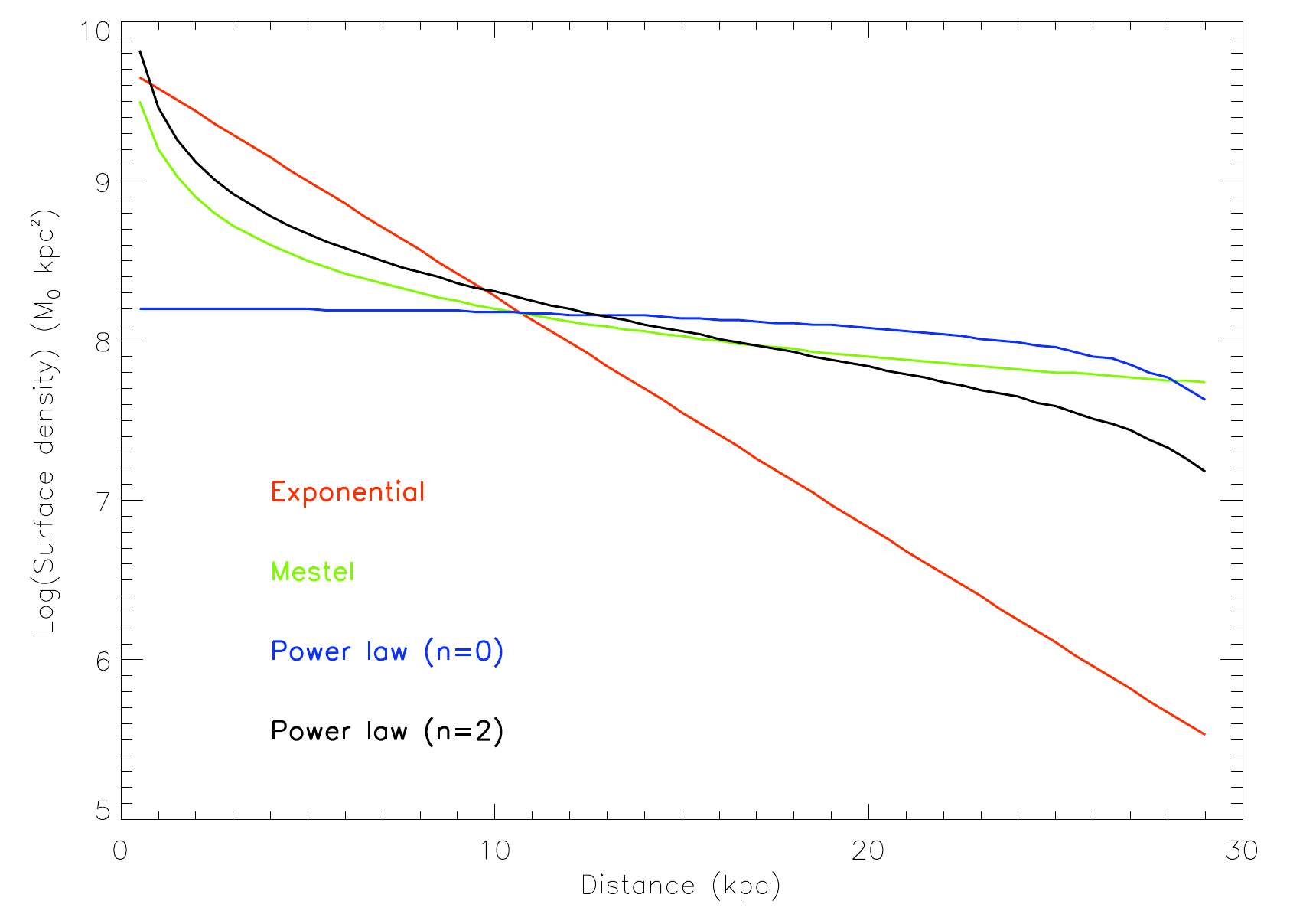}
\caption{The disc surface density profiles, calculated numerically, for the collapse of a constant density ($n=0$) (blue) and power law ($n=2$) (black) spherical cloud. 'Mestel disc' and exponential surface densities are shown in green and red respectively. Each profile is for a disc mass of $M=3\times10^{11}$ M$_{\odot}$ and radius of $R_{0}=30$ kpc, the exponential distribution has a scale length of 3 kpc.}
\end{figure}

As for the constant density spheroid case we can now use equations 1 and 2 to obtain the predicted rotation curve. Mestel's conclusion with regard to constant linear velocity was derived by simplifying a series expansion and assuming that the disc size tends to infinity. In Fig. 3 we show the result of the numerical solution of the predicted rotation of a finite disc both for the case where the disc surface density varies as $1/r$ and that obtained for the collapse of a sphere with density power law index $n=2$. Mestel's assertion of constant linear velocity is almost, but not quite, correct for the finite disc as the velocity increases from about 200 to 230 km s$^{-1}$ over the radius of the disc (Fig. 3 green line). The predicted Mestel constant disc velocity ($\sqrt{GM/R_{0}}$, see below) is actually about 211 km s$^{-1}$. The predicted velocity for the collapsed $n=2$ power law is similar to the $1/r$ case in that it produces almost constant velocity (Fig. 3 black line). Our conclusion is by no means new, but because of its simplicity it is worth repeating: we can achieve an approximately constant linear velocity disc from the collapse of a singular isothermal sphere. 

Before the advent of dark matter the dominant mass component of galaxies had been taken to be the stars. The challenge then was to model the rotation using essentially the stars alone. The stars were observed to follow an exponential surface brightness distribution (Freeman 1970) and so alone they could not account for the rotation. For completeness in Fig 3 and 4 we show the corresponding rotation curve and surface density for the exponential disc.

The specific cases of constant angular velocity and constant linear velocity are of particular interest for those interested in the origin of galaxy rotation curves. Ignoring the fine structure, many (all?) rotation curves can be modelled as combinations of constant angular and linear velocity components. Dwarf galaxies often have continuously rising rotation curves (constant angular velocity) while galaxies like the Milky Way have almost flat rotation curves (constant linear velocity) over most of the disc. Other galaxies have some combination of the two. Following Mestel's reasoning, the implication is that the original gas clouds that galaxies formed from, assuming that this was monolithic collapse have, to varying degrees, constant density cores surrounded by more diffuse isothermal spheres.

\begin{figure}
\centering
\includegraphics[scale=0.55]{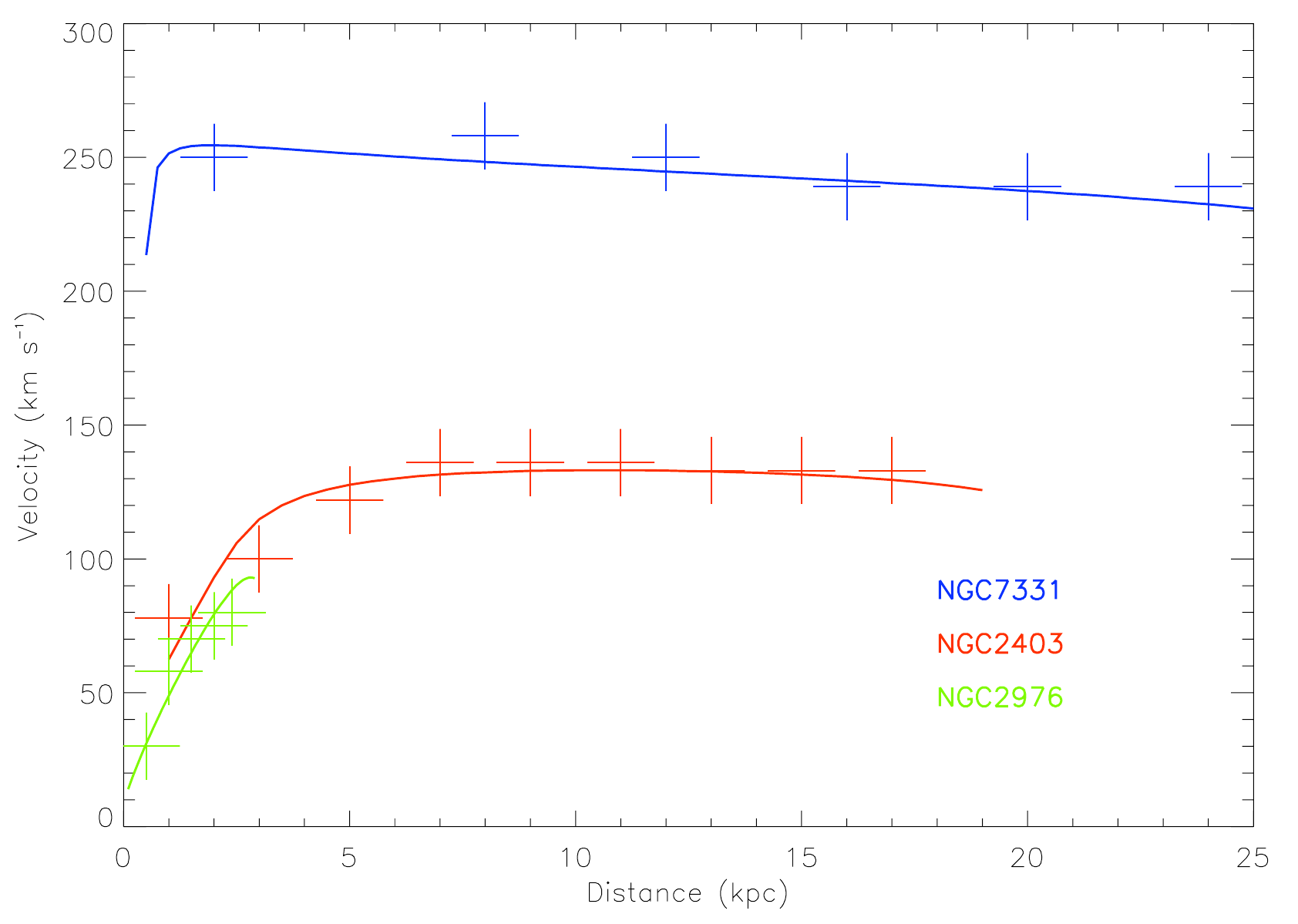}
\caption{The equilibrium rotation curves fitted to observational data for three galaxies taken from the THINGS survey. The origin of NGC7331 is almost entirely due to the collapse of a singular isothermal sphere gas cloud, NGC2403 requires a constant density core within the isothermal sphere while the origin of NGC2976 is a constant density cloud. }
\end{figure}

Let us now consider some specific cases. We have taken data points from the rotation curves of three galaxies studied as part of the THINGS survey (de Blok et al., 2008). The galaxies were chosen to illustrate the resultant rotation of discs that collapse from singular isothermal spheres, constant density spheres and some combination of the two. NGC7331 can be modelled as a singular isothermal spherical cloud ($\rho(w) \propto w^{-2}$) of mass $M=2.8\times10^{11}$ $M_{\odot}$ and size $R_{0}=30$ kpc. With this density law the cloud collapses to give a surface density distribution ($\Sigma(r)$) that is approximately that of the 'Mestel disc' ($\propto r^{-1}$) and hence an almost flat rotation curve (Fig. 5). NGC2976 is a much less massive galaxy and has a rising rotation curve. This can be modelled as resulting from the collapse of a $M=4.5\times10^{9}$ $M_{\odot}$ constant density cloud of size $R_{0}=3$ kpc (Fig. 5). NGC2403 requires a combination of a constant density core (size$\approx 2.5$ kpc) and $\rho(w) \propto w^{-2}$ outer regions. The total mass is $M=6.0\times10^{10}$ $M_{\odot}$ and size $R_{0}=20$ kpc, 14\% of the mass is contained within the constant density core (Fig. 5). 

The rotation curves of the above three galaxies can all be explained as resulting from the collapse of gas clouds with the most simple and plausible density distributions. Small clouds are of almost constant density while large clouds become more diffuse in their outer regions. As far as we are aware all galaxies have constant angular velocity as you approach their centres, which must arise from an initially approximately constant density core in the primeval cloud. The collapse of this constant density core leads to an almost constant surface density (Fig. 4). This measurement of a constant density  at the centres of galaxies (de Blok et al. 2003) has been a long standing thorn in the side of those who model galaxies using cold dark matter halos. Cold dark matter haloes are predicted to have steep `cuspy' density profiles (Navarro et al. 1996).

\section{Scaling relations}
If disc galaxies are primarily Mestel discs we can very easily predict what the fundamental scaling relations between their physical characteristics should be. It is clear from the above that the three quantities that define a Mestel disc are the radius ($R_{0}$), total mass (M) within $R_{0}$ and the velocity ($v_{m}$) at $R_{0}$.  Where for stability $v_{m}=\sqrt{\frac{GM}{R_{0}}}$, so just two of these three parameters are independent. We have used the extra-galactic database LEDA to extract data for a large (5174) sample of galaxies. Galaxies were selected such that they have angular diameters ($r_{25}$) greater than 1 arc min, an I band magnitude, a 21cm line flux, a measure of the maximum rotation velocity of the gas and finally that they have a morphological type of $t>2$ (Sab or later). 

Relatively large angular sized galaxies were chosen so that hopefully they will have been accurately measured, particularly their sizes. I band magnitudes and 21cm line fluxes were required to calculate stellar and gas masses. We can equate the LEDA `maximum rotation velocity of the gas' directly to $v_{m}$. Finally, by using $t>2$ we hopefully restrict our analysis to galaxies dominated by their disc. 

It is clear that we have independent measures of the velocity and size and so it is the mass of the galaxy that becomes the dependent variable. The only remaining problem is how does the size given in LEDA ($r_{25}$) equate to the size $R_{0}$ used in our model (Note: $R_{0}$ is the physical size of the disc), which we are assuming is the same as the cut-off observed at 21cm (see below). During their analysis of THINGS galaxies Portas et al. (2008) quote the measurements of Broeils and Rhee (1997), who give a value of $r_{HI}/r_{25}$=1.7 ($\pm 0.5$) measured at a column density of $1.25\times10^{20}$ cm$^{-2}$. This column density seems to agree very well with the cut-off seen in the surface density profiles of Fig. 2 and is the value we will use to scale the optical to the HI disc cut-off radius. Thus, we now have the two independent variables we require, $v_{m}$ and $R_{0}$, they are plotted against each other in Fig. 6.

\begin{figure}
\centering
\includegraphics[scale=0.55]{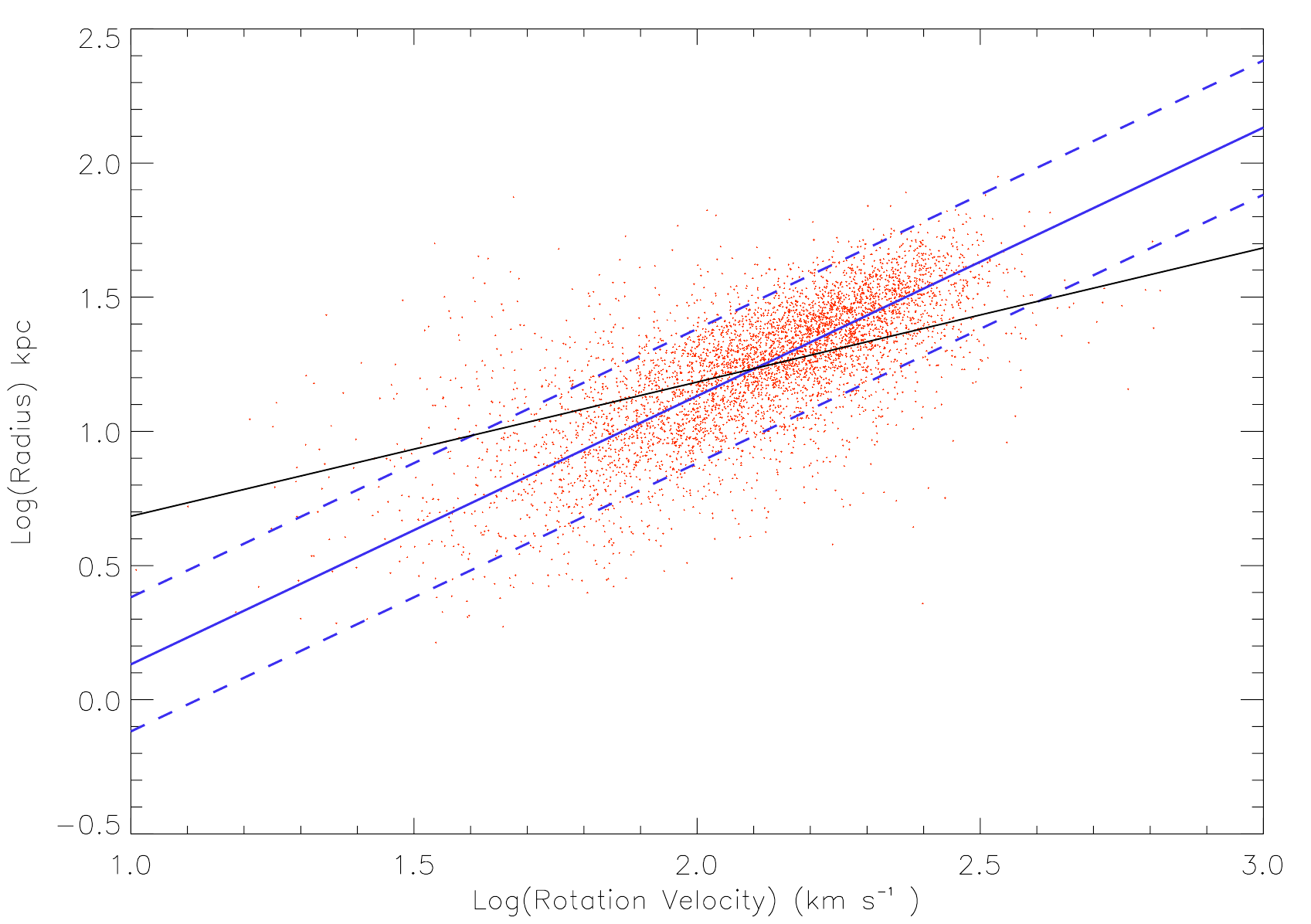}
\caption{The velocity/radius fundamental scaling relation using 5174 galaxies from the LEDA database. The solid blue line is the best fitting line of slope unity. The two dashed lines are for densities a factor of three lower (upper line) or higher (lower line) than the best fitting line. The black line is what is expected if galaxies are assembled directly as discs so that $v_{m} \propto R_{0}^{1/2}$.}
\end{figure}

Let us assume that isothermal spheres give rise to Mestel discs that produce the flat rotation curves we observe. If all galaxies initially form from similar density gas then there should be a mass size scaling relation $M \propto R_{0}^{3}$. Using the Mestel disc equation for velocity and the substitution for $M$ leads to the prediction $R_{0} \propto v_{m}$. Although true for Mestel discs this $v_{m}/R_{0}$ relation is not universally true for all disc surface densities. For example from Fig. 3 it is clear that the maximum rotation velocity of the exponential disc is much higher than that predicted by a Mestel disc or by the collapse of the uniform sphere. So it is important to note that the above scaling relations only apply to discs that result from the collapse of spheres with $0\le n \le 2$. On Fig. 6 we have drawn the best fitting line of slope unity and it is clear that it fits the data very well. 

However, the measured gradient of the best fitting line is 0.8 and not 1.0, but there is an obvious explanation of this discrepancy. Lines of slope unity in this plot correspond to lines of constant initial density in the proto-galactic spheroid ($\rho_{0}$) and so the fit is dependent on the relative numbers of galaxies of different densities along the line. The range of densities is actually remarkably small, the blue dashed lines on Fig. 6 illustrate a change in initial proto-galactic cloud density of only a factor of 3 from the mean, with most galaxies being constrained within these two dashed lines.

If galaxies were assembled directly as discs then we might expect a different relation between  mass and size i.e. $M \propto R_{0}^{2}$ and consequently that $R_{0} \propto v_{m}^{2}$ assuming in this case that the surface density remains constant. This relationship is also plotted on Fig. 6 (black line) and seems to be at odds with the data. It appears that galaxies retain a memory of their spheroidal origin.
If isothermal spheroids collapse to Mestel discs then we have $M=4\pi \rho_{0} R_{0}^{3}=2\pi \Sigma_{0} R_{0}^{2}$ a relationship ($\Sigma_{0}=2 \rho_{0} R_{0}$) between the measured surface density of the disc and the density of the cloud it formed from.

We can now define a 'typical' galaxy using the LEDA data. The mean maximum rotational velocity of the sample galaxies is $v_{m} \approx 144$ km s$^{-1}$, mean radius $R_{0} \approx 20$ kpc and resultant mean mass $M_{Rot}=1.6 \times 10^{11}$ M$_{\odot}$. The `characteristic' density ($\rho_{0}^{c}$) of the cloud that this `typical' galaxy formed from is $7 \times 10^{-23}$ kg m$^{3}$. If the current value of the baryonic mass density of the Universe is $\approx 2 \times 10^{-26}$ kg m$^{-3}$ (Spergel et al. 2007) then $\rho_{0}^{c}$ is achieved at a redshift of about 14 - not an unreasonable redshift for galaxy formation.

We now need to consider how much mass is required to be consistent with the observed velocities and sizes of our sample galaxy. The predicted (dynamical) mass of each galaxy is simply calculated using $M=\frac{v_{m}^{2} R_{0}}{G}$. Substituting for $v_{m}$ or $R_{0}$ in this equation leads to $R_{0} \propto M^{1/3}$ and $v_{m} \propto M^{1/3}$ (we also have $\Sigma_{0} \propto M^{1/3}$). We suggest that the latter relation ($v_{m} \propto M^{1/3}$) is the origin of the Tully-Fisher relation (see below). In Fig. 7 we show these two scaling relations for our LEDA sample. Again variations in initial density will produce scatter in the plots, but they are both extremely well fitted by a straight line with exponents of 0.33 and 0.34, for the velocity/mass and radius/mass relations respectively. This is remarkably close to expected value of 1/3 for Mestel discs.

\begin{figure}
\centering
\includegraphics[scale=0.5]{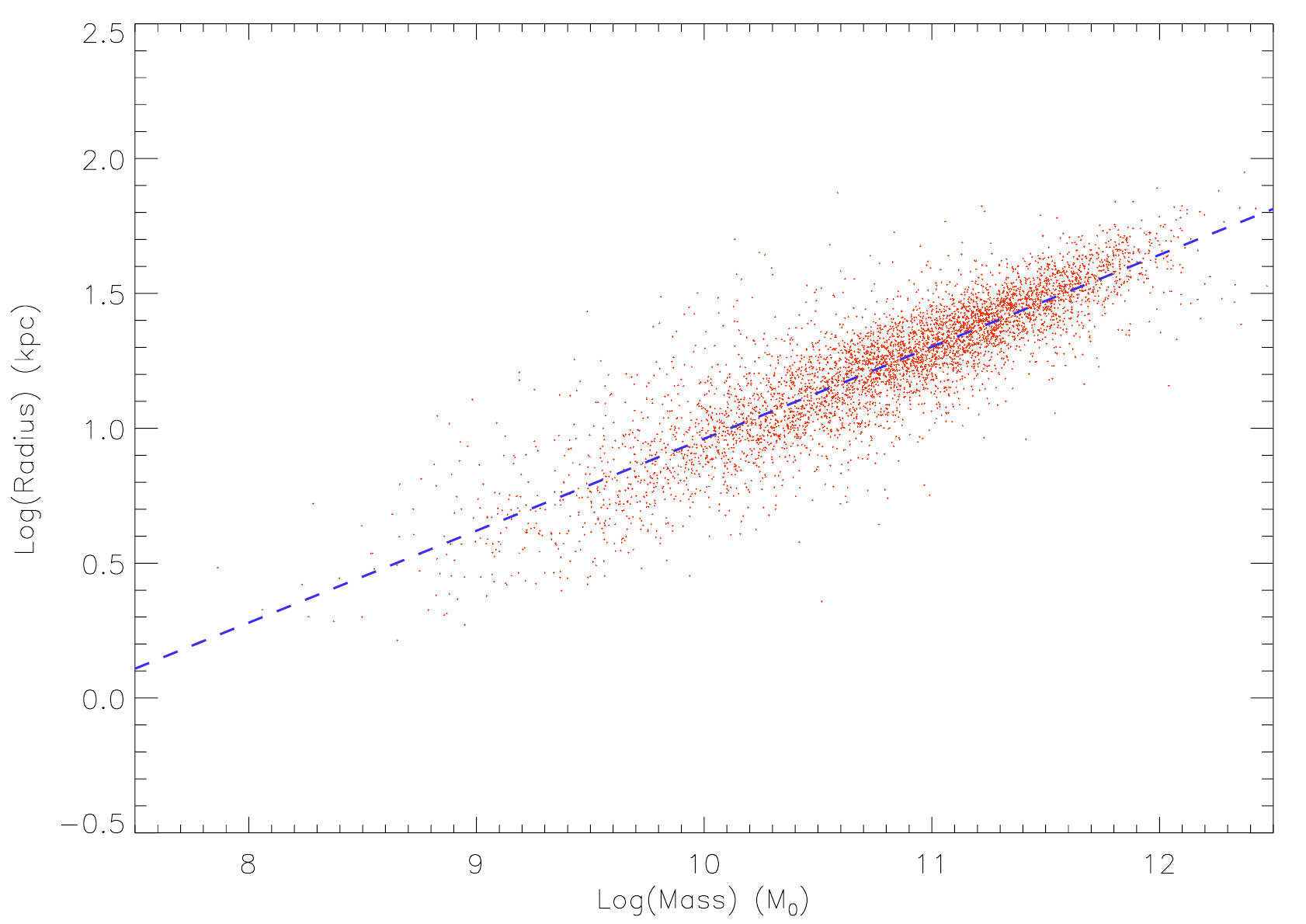}
\includegraphics[scale=0.5]{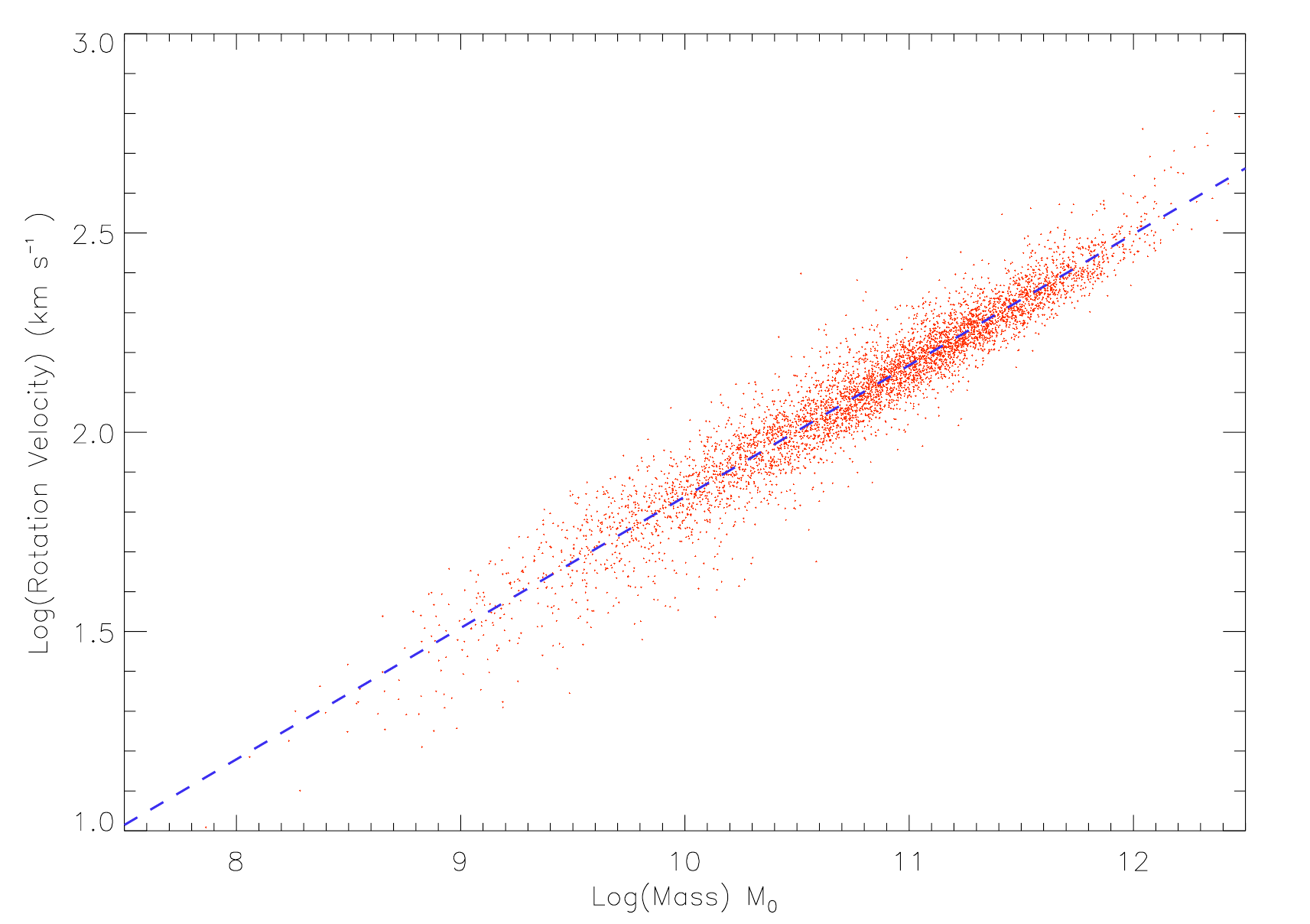}
\caption{The fundamental scaling relations of galaxies using 5174 galaxies from the LEDA database. Top - the mass size relation. Bottom - the mass maximum velocity relation. In each case the blue dashed line is a linear least squares fit to the data. }
\end{figure}

We have also used the LEDA I band magnitudes and 21cm line fluxes ($f$) to calculate the stellar and atomic gas masses of our sample galaxies. Stellar masses have been simply calculated assuming Sun like stars and $M_{I}^{\odot}=4.08$. Atomic gas masses are calculated using the standard formula $M_{HI}=2.4 \times 10^{5} d_{Mpc}^{2} \int fdv$. In each case we have used the LEDA recession velocities and a value of H$_{0} = 72$ km s$^{-1}$ Mpc$^{-1}$ to get distances. With a mean stellar mass of $1.8 \times 10^{10}$ M$_{\odot}$ and atomic gas mass of $0.7 \times 10^{10}$ M$_{\odot}$ there is not a huge discrepancy between the baryonic mass we can account for and the mass required for rotation ($<M_{Rot}>=1.6 \times 10^{11}$ M$_{\odot}$). The mean ratio of dynamical mass to stellar and atomic gas mass for the whole sample is $\approx 7$ (Fig. 8). This scaling factor of 7 is consistent with the cosmological 'missing mass' problem which measures $\Omega_{Matter} \approx 0.24$ and $\Omega_{Baryon} \approx 0.04$ leading to $\frac{\Omega_{Matter}}{\Omega_{Baryon}}=6$ (Spergel et al., 2007).

As we will see in the following sections we will suggest that the mass discrepancy between what we can `see' ($M_{Star}+M_{HI}$) and that measured via rotation $(M_{Rot})$, is `hidden' in a previously undetected gas component. To assess just how much gas is 'missing' we also calculate the factor $k$ which multiplies the atomic gas mass to fully account for the dynamical mass i.e. $k=\frac{M_{Rot}-M_{Star}}{M_{HI}}$, its mean value for the LEDA sample is 27 and its distribution of values is also shown in Fig. 8. 

Given the gas disc scaling relation (Bosma 1978, 1981) and that the observed surface density of stars appears to be exponential we require an additional mass component, larger than that of the stars, but distributed like the gas to account for the rotation. As a demonstration that models like this are possible and stable Kalberla and Kerp (1998) demonstrate that a thick ($\approx 4$ kpc) disc of mass $\approx 5 \times 10^{11}$ M$_{\odot}$ can account for the rotation of the Milky Way (see also Kalberla, 2003, 2004, Kalberla et al. 2007 and Revaz et al. 2009).

In the following sections we will consider observations of the mass of gas in galaxies, how uncertain these might be and whether this uncertainty can account for the value of $k$ and ultimately the dynamical mass of galaxies. 

\begin{figure}
\centering
\includegraphics[scale=0.55]{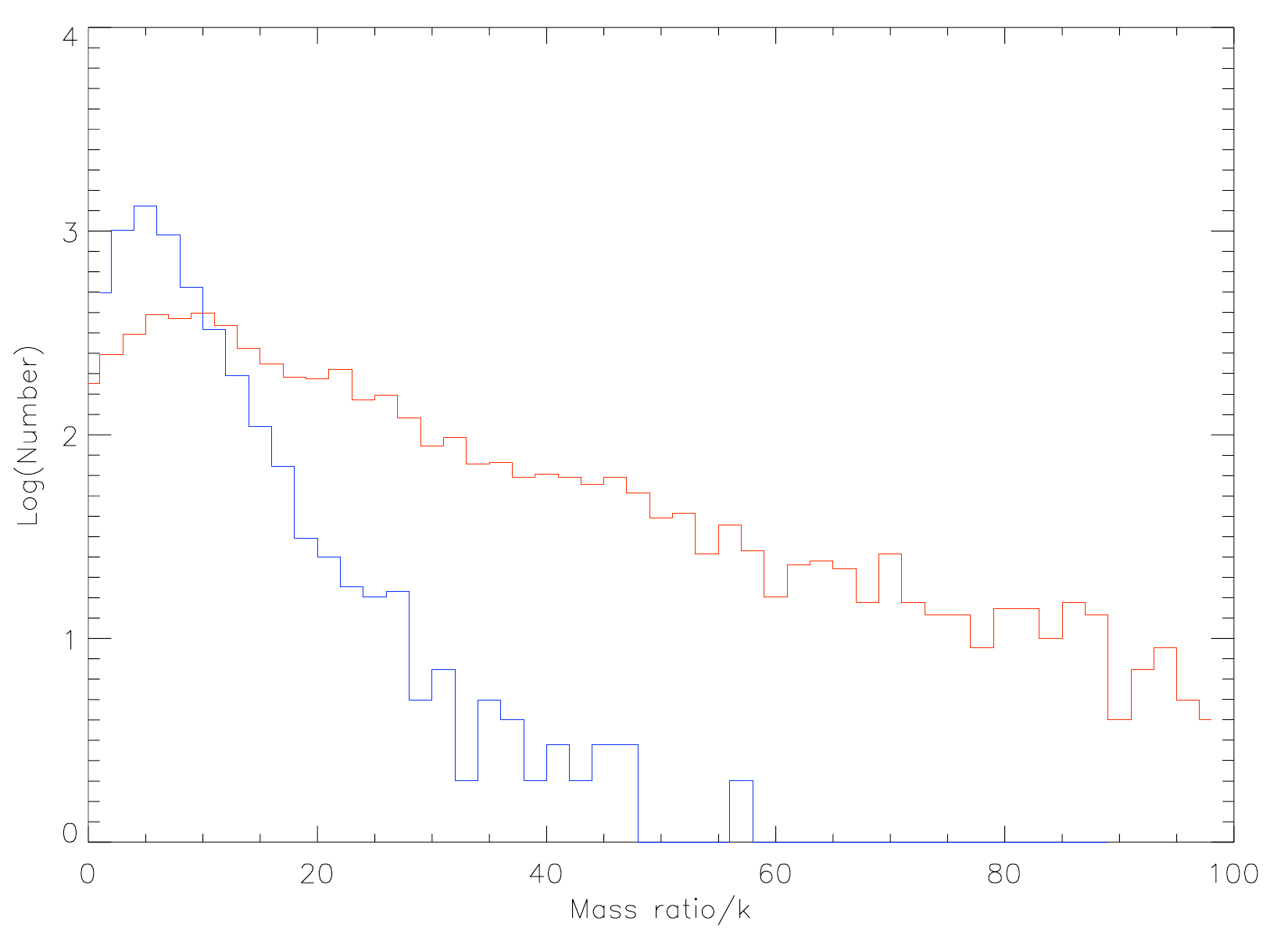}
\caption{The blue histogram shows the distribution of values of the ratio of, dynamical mass to mass in stars and atomic gas. The red histogram shows the distribution of the value of $k$ that is needed to multiply the atomic gas mass by so that the rotation can be accounted for.}
\end{figure}

\section{Observation of gas in galaxies}
\subsection{Atomic hydrogen}
 As noted above the THINGS team measure an approximately constant atomic hydrogen surface density for galaxies (Bigiel et al., 2010). 
The value of the observed HI column density ($10^{20}<N<10^{21}$ cm$^{-2}$) is intriguing. The optical depth of atomic hydrogen (Kerr, 1975) is:
\begin{equation}
\tau=\frac{5.5\times10^{-19}N}{T_{S}\sigma}
\end{equation}
For $T_{S}=100$ K , $\sigma=10$ km s$^{-1}$ clouds with column densities ($N$) of about $10^{21}$ cm$^{-2}$ or above are opaque. $\tau$ is inversely proportional to $T$ and $T_{S}=$100K is an upper limit on the cool atomic gas temperature measured by Dickey et al. (2009). Almost all inferences made about HI in galaxies assume that it is optically thin.

Ionisation of the HI will occur (re-combinations=ionisations) when $N\approx\frac{2\phi}{\alpha n}$ where $\phi \approx 10^{4}$ photons cm$^{-2}$ s$^{-1}$ is the intensity of the ionising background and $\alpha=3\times10^{-16}T_{S}^{0.75}$ cm$^{3}$ s$^{-1}$ is the hydrogen recombination coefficient (Maloney, 1993). For a disc with a gas $z$ scale height of $\beta$ the above equation becomes: $N \approx \sqrt{\frac{4\phi\beta}{\alpha}}$. For a gas scale height of $\beta=500$ pc and again a temperature of 100 K, the column density at which ionisation becomes important is $N\approx10^{19}$ cm$^{-2}$. At column densities lower than this there is an increasing ionisation fraction. 
The typical explanation of the sharp cut-off to the HI profiles shown in Fig. 2 is that this is due to the ionising background. But, as noted by Portas et al. (2008), the drop in column density starts at $\approx10^{20}$ cm$^{-2}$, which is too high for the HI to be substantially affected by ionisation. Hence our suggestion that this really does mark the edge of the disc.

The predicted range of observed column densities due to ionisation and optical depth ($10^{19}<N<10^{21}$ cm$^{-2}$) and the THINGS result for other galaxies ($10^{20}<N<10^{21}$ cm$^{-2}$) can be compared with observations of our Galaxy. Stark et al. (1992) made a map of the HI column density across the sky and as expected the measured column densities are somewhat restricted (Fig. 9). There is quite a sharp cut-off at $N=10^{20}$ cm$^{-2}$ and then it just slowly falls away to 10$^{22}$ cm$^{-2}$. Most lines of sight lie in the predicted range 10$^{20-21}$ cm$^{-2}$.  The Stark et al. survey column density limit is $\approx 2-3\times10^{19}$ cm$^{-2}$. Thus the cut-off at low column densities (Fig. 9) appears to be real rather than an artefact of the observational set up. 
One interpretation of the narrow range of 21cm column densities observed for both our Galaxy and THINGS galaxies is that the highest column density sight lines are optically thick and hiding substantial amounts of atomic gas. 

An alternative way of measuring the HI column density along a line of sight, that does not rely on 21cm emission, is to use stellar spectra. Abundant atomic hydrogen should cause strong absorption particularly in the Lyman $\alpha$ line. Bohlin et al (1978) have mapped the HI distribution using Lyman $\alpha$ absorption lines seen in the spectra of 100 stars observed by the Copernicus satellite. Interpretation of these lines is not straight forward because the limiting distance for these ultra-violet observations is dependent on the dust extinction, which is itself dependent on the HI column density. So the densest gas regions are un-observable. Bohlin et al. placed a limit of E(B-V)$<0.5$ for sight lines to the stars they observed, which translates, into a HI column density of less than $2 \times 10^{21}$ cm$^{-2}$. The distribution of column densities sampled by the stars is also shown in Fig 9.,  where it is compared to that derived from the 21cm emission data. It is clear that the distributions are very different making it difficult to say if either is a true reflection of the real atomic column densities through the disc. What is interesting is that the stellar data has increasing numbers of sight lines with increasing column density while the HI data has decreasing numbers with increasing column density. One can speculate whether this trend for the stellar data would continue, and hence make the inference of much more atomic gas, if it was not arbitrarily cut-off at just over 10$^{21}$ cm$^{-2}$.  Bolin et al. also, rather interestingly, point out that in a number of cases the 21cm and Lyman $\alpha$ measurements do not agree on the measured column density. If optically thin the 21 cm observations should sample a longer HI column than the distance to the star so $N_{\mbox{Ly}_{\alpha}} \le N_{\mbox{21cm}}$, but this was not always the case.

\begin{figure}
\begin{center}
\includegraphics[scale=0.55]{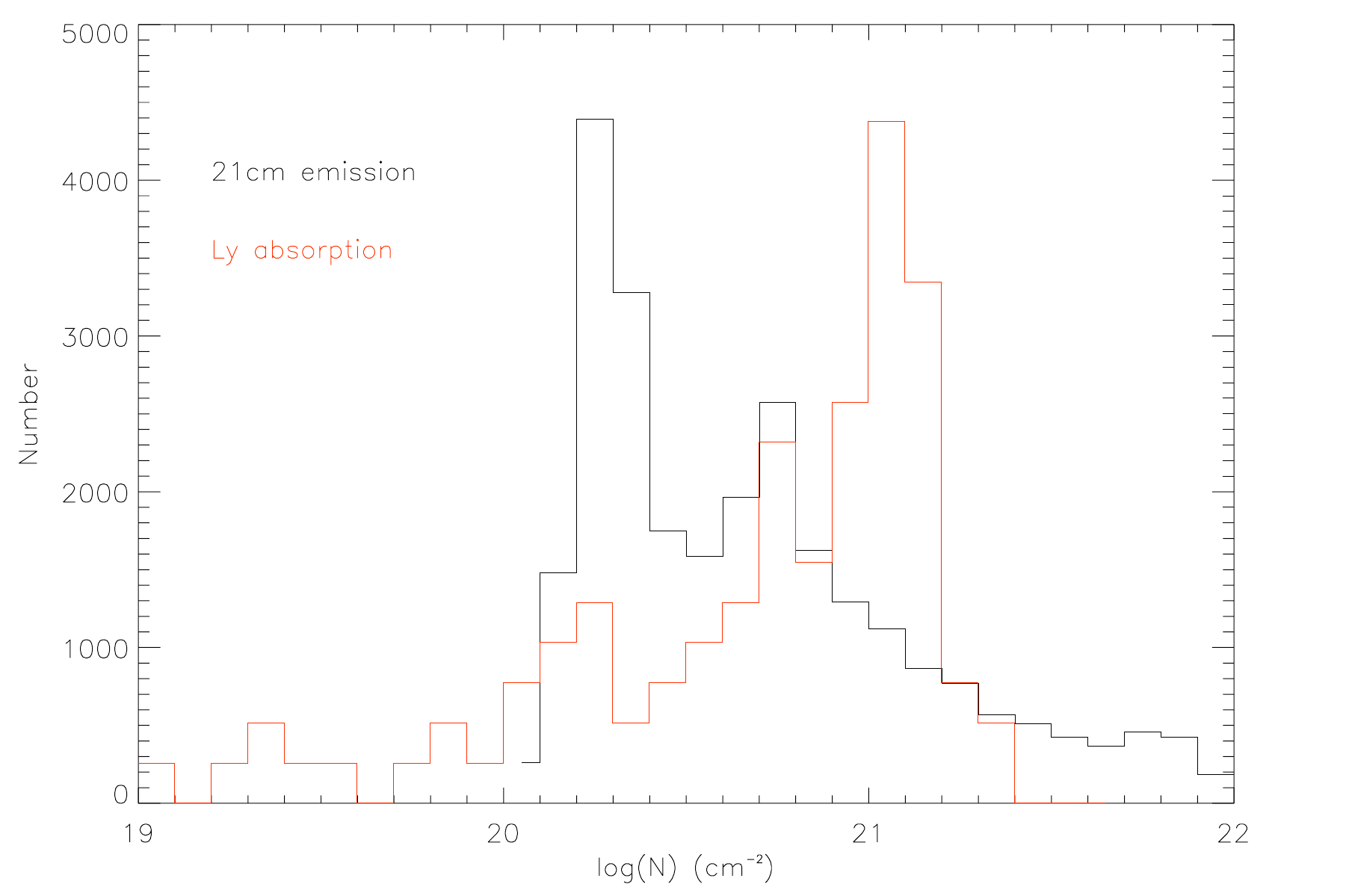}
\caption{The distribution of atomic hydrogen column densities from the Stark et al. (1992) observations of our galaxy (black). The red line shows the column densities derived from ultra-violet observations of nearby stars (scaled to give the same peak value).}
\end{center}
\end{figure}

How much atomic gas could there be in galactic discs without contradicting the Stark et al. and THINGS observations? Let's return to our 'typical' Mestel disc. In his paper Mestel (1963) also solves the problem of the predicted surface density distribution of a truncated disc that has constant linear velocity finding that:
\large
\begin{center}
$\Sigma(r)=\frac{v^{2}_{m}}{2 \pi Gr} \left[1-\frac{2}{\pi}\sin^{-1}\left(\frac{r}{R_{0}}\right) \right]$
\end{center}
\normalsize
This now has zero column density at and beyond the truncation radius $R_{0}$. We will use this truncated disc to model the opacity effect and what it predicts.

By requiring that $v_{m}$=211 km s$^{-1}$ and that $R_{0}$=30 kpc for our `typical' galaxy we have a total disc mass of $M=3\times10^{11}$ $M_{\odot}$, which we now assume is all in the form of atomic hydrogen distributed to an exponential scale height of 0.5 kpc (Bohlin et al. 1978). We have constructed a simple numerical radiative transfer model of a Mestel gas disc using the above parameters, including the effects of 21cm opacity (eq. 5 above) so that we can compare with observations. The predicted observed (face-on) radial 21cm column density distribution is shown in Fig 10 along with the actual column density.  The predicted almost constant observed column density has about the correct value of $10^{21}$ cm$^{-2}$ (compare with Fig 2) while the real column density is more than an order of magnitude higher over most of the disc. This HI disc, massive enough to account for its rotation, has an almost constant observed surface density of hydrogen consistent with observation. There is however a `catch'. To achieve the necessary opacity we had to use a gas temperature of 10K (and $\sigma=$10 km s$^{-1}$). Thus to be able to account for the rotation using atomic hydrogen alone we need it to be cold. Can cold atomic gas be present in galactic discs? There are three types of observation that provide us with information on the physical state of the atomic gas.

\begin{figure}
\begin{center}
\includegraphics[scale=0.55]{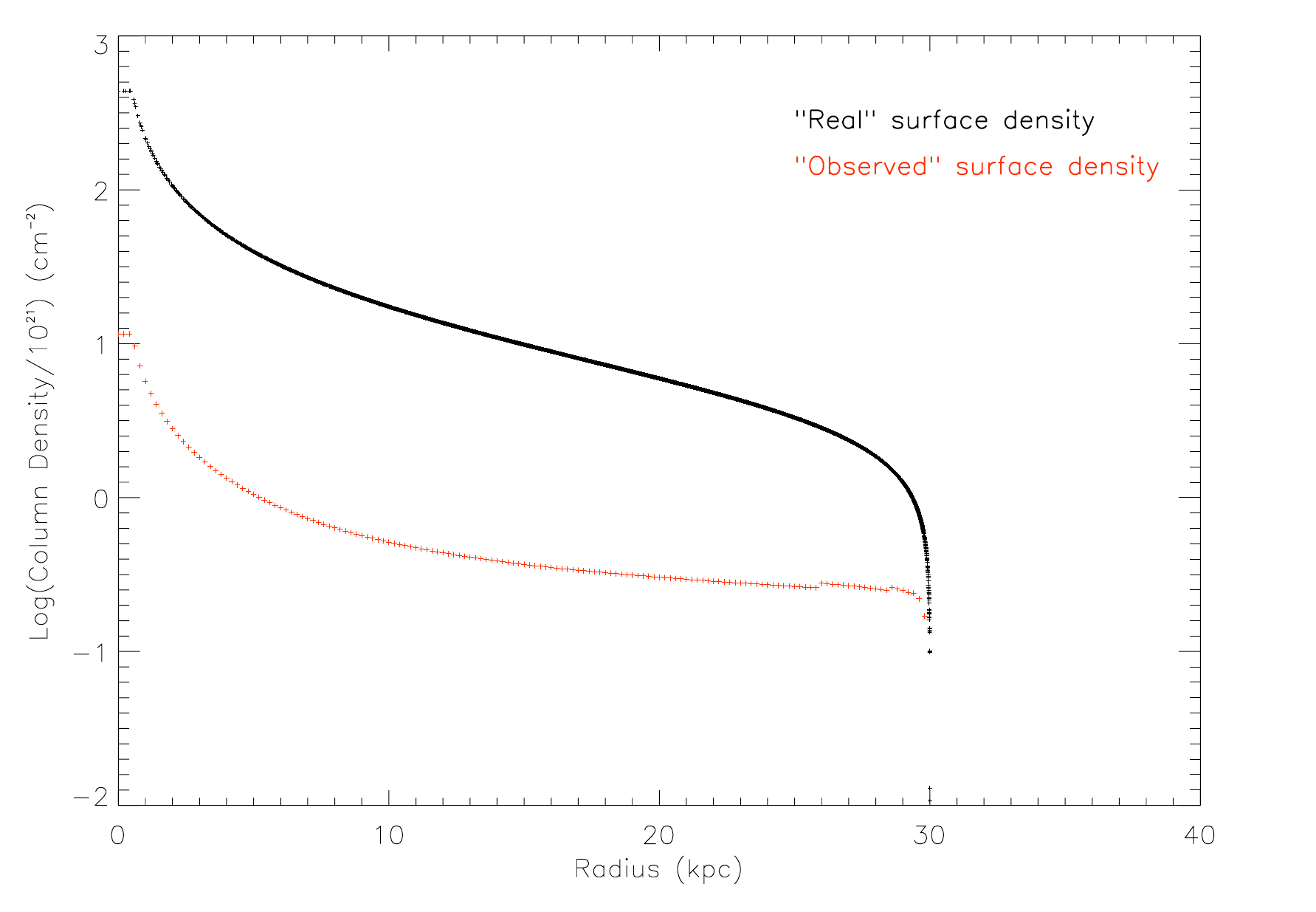}
\caption{The real (black) and observed (red) atomic hydrogen column density profile of an optically thick face-on Mestel disk. In this case the temperature of the gas is 10K, it has a velocity dispersion of 10 km s$^{-1}$ and a $z$ scale height of 0.5 kpc. Note for comparison with Fig. 4 that $10^{21}$ cm$^{-2} \approx 8 \times 10^{9}$ M$_{\odot}$ kpc$^{-2}$. }
\end{center}
\end{figure}

The first and most familiar is the emission line surveys such as THINGS. The difficulty with these is that the emission is independent of the gas temperature when it is optically thin and proportional to the temperature when optically thick. Thus observations of the emission do not lead directly to a measurement of the column density unless the gas is assumed to be optically thin, which is what is almost universally done. 
Returning to the Stark et al. observations of the distribution of HI column densities on the sky (Fig. 9) these can actually be interpreted in two ways. Previously the HI has been assumed to be optically thin so the observed distribution of column densities would be regarded as a true measure of the column densities along each line of sight through the Galaxy. This is not so if these sight lines are optically thick. Using our above radiative transfer model of a typical galaxy we can predict what we would observe from about the position of the Sun in the mid-plane. For gas with $T\sigma=100$ K km s$^{-1}$ the observed column density would be about 10$^{20.5}$ cm$^{-2}$ in all directions independent of the actual column density. For $T\sigma=1000$ K km s$^{-1}$ the observed column density increases to about 10$^{21.3}$ cm$^{-2}$. Thus an alternative interpretation of the Stark et al 21cm data is that it is really a measure of $T\sigma$ along the line of sight in the optically thick regime and large amounts of atomic gas could be hidden there.

Secondly, 21cm absorption studies generally target background continuum radio sources so that absorption by hydrogen in our galaxy can be studied (Dickey et al. 2009, Heiles and Troland, 2003, Kolpack et al. 2002, Garwood and Dicky, 1989, Dicky et al. 1978). These observations are primarily concerned with absorption `features' and as such they rely on the premise that the hydrogen is contained within discrete clouds - it is clearly much more difficult to detect hydrogen extended both spatially and in velocity. The surveys do provide useful physical parameters for the gas in clouds, which is normally divided into cold and warm components. Typically the cold component is observed to have a spin (kinetic?) temperature of about 50K with some less than 25K, while the warm component has $T>500$K. Absorption by the warm component is generally at larger distances above the Galactic plane than the cold component. Interpretation of the absorption lines is fraught with difficulties: lines may be blends from different clouds, the on source absorption spectrum may contain a contribution from HI emission in the beam, baselines are difficult to measure and to measure optical depths you need an emission measure that is determined off-source. Data is also often difficult to interpret because authors give properties of the clouds, but not the integrated optical depth along the line of sight. It is the cold component (eq. 5) that gives rise to high optical depths, the warm component's optical depth being generally pretty low. 

An exensive study of 21cm continuum source absorption through the gas disc of our galaxy was carried out by Kolpak et al. (2002). Their stated sensitivity gives a minimum detectable column density of $7\times10^{19}$ cm$^{-2}$ though this is in `clouds' and so provides only an estimate of the minimum column density (and opacity) along a line of sight. Kolpak et al.'s objective was to calculate a radial 21cm opacity ($\tau$) profile for the Galaxy. They reach four important conclusions. Firstly, that within 3-8 kpc from the Galactic centre (they have few sight lines within 3 kpc) the HI opacity lies between  $\tau=$0.8 and 1.5 with some sight lines having opacities as high as 3.5. The inferred column densities of hydrogen depend on the temperature and velocity dispersion along the line of sight, but these observed opacities are not at all inconsistent with our optically thick `typical' galaxy. Kolpak et al. say that there are saturated lines at all radii and presumably these are given a minimum $\tau$ consistent with this saturation. Secondly, low opacities are measured beyond 8 kpc, but at these radii they say opacities are very uncertain because of the uncertainties in the rotation curve. Thirdly, there is a good correlation of high HI opacity with high molecular gas emission, as might be expected if cold HI eventually leads to the formation of molecular clouds. Fourthly, there is quite strong disagreement about the opacity distribution compared to a similar study carried out by Garwood and Dickey, (1989), which further illustrates the difficulty of interpreting absorption line studies.

The third and final way of measuring the quantity of neutral hydrogen is by HI self-absorption. HI self-absorption occurs when there is gas with a low excitation temperature lying infront of a hotter cloud. This of course requires a rather special geometrical alignment, but in the plane of the galaxy this seems to happen quite often (Gibson et al., 2005). Strong HI self-absorption again requires significant density enhancements (clouds) and a large scale background at the right velocity and/or very cold gas. This self-absorption does not provide a strong constraint on the total amount of atomic hydrogen present, but Gibson et al., (2005) do say, after carrying out an extensive self-absorption study of the Galactic plane that '...faint HI self-absorption is detected virtually everywhere that the HI emission background is sufficiently bright'. 

There are of course quite different problems associated with discerning the mass of atomic hydrogen in our Galaxy compared to other galaxies because we reside amongst the hydrogen. Our current understanding of the Milky Way has benefited greatly from observations made of other galaxies, particularly our nearest neighbour M31. Recently Braun et al. (2009) have found what they describe as `ubiquitous HI self-opacity features' over the face of the disc of M31. They say that in some cases there is an order of magnitude difference in the HI column density inferred using the normal optically thin assumption, but that currently they account for only about a 30\% increase in the previously determined HI mass. This they say though must be a lower limit because of the geometry requirements of HI self-absorption stated above and the blending of lines. They find gas spin temperatures of 20-60 K for the opaque gas, consistent with those required to give high opacity in our `typical' model galaxy. It is also worth noting that they find a strong truncation of the HI disc at 28 kpc (at $5\times10^{19}$ cm$^{-2}$).

Perhaps the most convincing observation consistent with the idea that there is both an extensive and massive cold HI component to galaxies comes from the observations made by Knee and Brunt (2001). We quote from the abstract of their paper - 'Cold HI is difficult to observe, and therefore our knowledge of its abundance and distribution in the interstellar medium is poor. The few known clouds of cold HI are much smaller in size and mass than typical molecular clouds. Here we report the discovery that the HI supershell GSH139-03-69 is very cold (10K). It is about 2 kpc in size and as massive as the largest molecular complexes. The existence of such an immense structure composed of cold atomic hydrogen in the interstellar medium runs counter to the prevailing view that cold gas resides almost exclusively in clouds dominated by molecular hydrogen'.

\subsection{Molecular hydrogen}
Another possible substantial component of the interstellar medium in the discs of galaxies is molecular gas, primarily in the form of molecular hydrogen. The direct observation of molecular hydrogen presents substantial problems because it has no permanent dipole, so there is no pure rotational spectrum, which for molecules would normally produce emission in the ground based mm window. $H_{2}$ rovibrational lines can be detected at wavelengths of a few micron, but these arise from gas at a few 1000K and so it does not correspond to the cold molecular material (10-40K) that is known to reside in the discs of galaxies. Generally the quantity of molecular hydrogen has been inferred from the relatively abundant CO molecule that does have a dipole moment and so emits strongly (Dickman 1978). 

The exact relation between the abundance of H$_{2}$ and CO has been a bone of contention over the years, with a current best buy value of $X_{CO} \approx 2.0 \times 10^{20}$ cm$^{-2}$ K$^{-1}$ km$^{-1}$ s where $X_{CO}=N(H_{2})/I_{CO}$, $N(H_{2})$ is the molecular hydrogen column density and $I_{CO}$ is the brightness temperature of the $J=1-0$ emission line at 2.6mm (Burgh et al. 2007).
Historically confidence in both the constancy of $X_{CO}$ and its precise value comes from three independent measures that concur (Solomon and Barrett, 1991):
\begin{enumerate}
\item The correlation of CO intensity with extinction ($A_{V}$) determined by star counts.  The $H_{2}$ column density is then inferred from the $N_{H}/A_{V}$ relation.
\item By modelling the $\gamma$ ray flux from the interaction of cosmic rays with hydrogen molecules and comparing to the CO intensity. 
\item Virial analysis using cloud sizes and line widths to get the total molecular mass and then comparing this mass to the CO intensity.
\end{enumerate}

Contrary to the above there are two quite logical reasons why we might not expect a good correlation between CO and $H_{2}$ and over more recent years these have been investigated. Firstly, the quantity of CO compared to hydrogen should depend on how metal enriched the gas is and so $X_{CO}$ should depend on metalicity. Estimates of the dependence range from $X_{CO} \propto Z^{-2.5}$ (Israel 2000) to $X_{CO} \propto Z^{-1}$ (Wilson 1995, Boselli et al. 2002). Given the metalicity gradients found in galaxies (Magrini et al. 2011) it is easy to see how the former ($\propto Z^{-2.5}$) metalicity dependence leads to the conclusion that $X_{CO}$ could be a factor of 10 higher in the outskirts of galaxies. For example using a sample of 29 SINGS galaxies Draine et al. (2007) have derived both HI and $H_{2}$ masses. Using their values we obtain a mean value of $<k>=(M_{HI}+M_{H_{2}})/M_{HI}=3.2$. This is based on a constant value of $X_{CO}$, which if made 10 times larger would give values of $<k>$ just about that required to account for galactic rotation
\footnote {Drain et al. (2007) use a value of $X_{CO} = 4.0 \times 10^{20}$ cm$^{-2}$ K$^{-1}$ km$^{-1}$ s.}.
In support of the above are observations of CO absorption in cold ($\approx 10$K) clouds in our galaxy at a radial distance of about 12 kpc (Lequeux et al. 1993). At this distance Lequeux et al. say that the mass of $H_{2}$ is already 4-5 times large than HI ($k=5-6$) and that it will only increase with increasing distance.

Secondly, models indicate that the photodissociation  of CO is more sensitive than $H_{2}$ to the strength of the interstellar UV radiation field (van Dishoeck and Black 1988, Kopp et al. 2000). Thus in different galactic environments we might expect different ratios of CO to $H_{2}$. Recently Ade et al. (2011) using PLANCK data have used just such an explanation to interpret an observed breakdown in the correlation between optical depth and hydrogen column density (between about $8 \times 10^{20}$ and $3 \times 10^{21}$ cm$^{-2}$). They suggest that the CO is being photodissociated and so it no longer becomes a good tracer of $H_{2}$. They describe the unaccounted for gas as 'dark' and infer that it amounts to an increase in the atomic gas by 28\% (though due to opacity effects this could be as much as 50\%) and that the molecular hydrogen mass could double. On their own these values are not large enough to account for galactic rotation, but when combined with the uncertainties due to the low metalicty of gas in the outer regions of galaxies they become interesting. 

Although the emission from cold ($T<100$ K) molecular hydrogen is rather weak there has been a detection of emission lines from it. Using ISO Valentijn and van der Werf (1999) detected $H_{2}$ emission at 17 and 28$\mu$m along the plane of the edge on galaxy NGC891. This emission they assign to a warm (130-200K) and colder (80-90K) molecular component, with the colder component becoming more dominant at larger distance from the galaxy's centre. They conclude that the $H_{2}/HI$ ratio is 3 in the inner regions, but that this rises to greater than 20 in the outer regions ($k=21$). Again this is very close to the value required to account for galactic rotation. 

So, if recent measures of molecular hydrogen indicate that there is very much more of it around than previously thought, what about the convincing three independent measures of $X_{CO}$ described above? The first general comment is that the universally constant $H_{2}$ CO relation is almost certainly far too simplistic possibly because all three measures were derived using things quite local to the Sun. Considering each measure in turn:
\begin{enumerate}
\item There is an assumption here of a fixed gas-to-dust ratio throughout the galaxy, but it probably changes with the metalicity gradient (Magrini et al. 2011).
\item The interpretation of the $\gamma$ ray data is very model dependent and critically dependent on both the local cosmic ray flux and gas density and how both of these change throughout the galaxy. Pfenniger et al. (1994) point out that the exponential scale length of $\gamma$ ray emission from the Galaxy is about three times longer than that of the sources that produce it (cosmic rays from pulsars and massive stars?). They suggest that the $\gamma$ rays are actually tracing `..some hidden form of hydrogen'. This `$\gamma$-ray problem' has not gone away being recently confirmed by Ackermann et al. (2011). Strong et al. (2004) suggest that the solution to this problem is that the value of $X_{CO}$ should be increased by a factor of from 5 to 10 when going from the inner to outer parts of the Galaxy. 
\item Molecular masses derived from the virial theorem are of course only applicable to gravitationally bound clouds. A problem for our cold gas `idea' for the `missing'  mass is why does it not form stars? We do not have a definitive answer for this except that it maybe a threshold effect (Kennicutt 1989), but the threshold is just a bit higher than previously thought. If the gas is not forming stars then it is probably not in the form of gravitationally bound clouds and the logic of this method of obtaining H$_{2}$ from CO no longer applies.
\end{enumerate}

\subsection{Comments on the gas mass of galaxies}
From the above discussion we believe that it is entirely plausible that galaxies contain significantly more gas than has been previously thought.  This plausible gas could easily have sufficient mass ($k\approx 27$) to account for the rotation of galaxies. Both cold atomic and molecular gas can be easily hidden in a galaxy and the observation of cold atomic hydrogen in our galaxy adds weight to our `missing' or `dark' gas hypothesis (Knee and Brunt 2001). 

Some comments on the formation of one form of hydrogen from the other may put things in perspective. As stated earlier the THINGS team assume that the almost constant observed column density of HI across the discs of galaxies is due to the formation of molecular hydrogen. Molecular hydrogen is thought to form by surface recombination on dust grains. The process depends on the flux of atomic hydrogen, the number density and cross-section of the grains and a poorly known `H$_{2}$-formation efficiency factor'. Given the observed metalicity and corresponding gas-to-dust gradients (Magrini et al. 2011) across galaxies one might infer that molecular gas formation is suppressed in the outer regions of galaxies and hence here is the realm of the atom. 

$H_{2}$ is also destroyed by photodisassociation in the interstellar medium unless it is self-shielded in dense clouds. Allen (2002) discuss an alternative view of these processes. He suggests that hydrogen in the discs of galaxies is primarily in the molecular form and that the atomic hydrogen we detect through the 21cm line is almost entirely the consequence of this photodisassociation. He goes on to model this process showing that the almost constant observed column density of atomic hydrogen is successfully explained in this way. This is a `chicken and egg' situation - which came first the atomic or the molecular form?  

We now have three ways of explaining the almost constant surface density of atomic hydrogen, it is either; the result of the onset of molecule formation, an opacity effect and/or the result of the disassociation of molecules. These three possibilities reflect our uncertainty in the relation between atomic and molecular hydrogen in galaxies and hence the possibility that more of both exist.

Finally, can big bang nucleosynthesis (BBNS) constraints on the baryon density provide sufficient material to account for the rotation of galaxies? Pagel (1999) provides a nice summary of the baryon accounting and its relation to BBNS. From BBNS he constrains the density of baryons to be $0.035h^{-2}_{70}$. Quoting Fukugita et al. (1998) Pagel gives a density in stars of $0.0035h^{-2}_{70}$ a factor of 10 lower than BBNS. When hot gas in clusters is included this raises the `detectable' baryon density to $0.0061h^{-2}_{70}$, now a factor of  about 6 different from BBNS. This discrepancy between observed and predicted baryons is typically accounted for by invoking a huge reservoir of baryons in the warm inter-galactic medium (see the introduction of Takei et al. 2011 for a recent discussion of the issue). An alternative is that they actually reside in the disc of galaxies. In section 3 we showed that the observational data requires a factor of about 7 increase in the baryonic mass of galaxies to be consistent with their dynamics, which we can now see is in good agreement with the early universe constraints on the baryon mass density. This additional `puzzle' should be added to our list in the introduction: `The fractional increase in the mass of a galaxy required to account for its rotation is just that needed to account for the discrepancy between baryons observed and those inferred from big bang nucleosynthesis'.

\section{Other relevant issues}

\subsection{An exponential disc of stars}
Just why the stars in a disc galaxy should have an exponential surface brightness distribution (Freeman, 1970) has been a problem for both those who assume disc galaxies form by monolithic gravitational collapse (Seiden et al. 1984) and those who more recently accept their hierarchical assembly (Dutton, 2009). In his seminal paper Freeman (1970) says that the answer to the exponential nature of discs is not known, but quoting Toomre (1964) he suggests that their origin may be something to do with the radial velocity dispersion of the stars, which needs to exceed a critical value to provide stability.
Seiden et al. (1984) argue that a power law (Mestel?) disc is the most likely form of the gas disc and that an `exponential like' surface density distribution for the stars comes about from the star formation prescription they use - Stochastic Star Formation (SSF). Although SSF may offer a satisfactory explanation of the way stars form in irregular, particularly dwarf irregular, galaxies it is not so clear that it can explain star formation in galaxies like the Milky Way. It is quite clear that star formation in spiral galaxies is concentrated along the spiral arms and so the distribution of stars must logically be related to the nature of the spiral density wave. Dutton's (2009) paper highlights the fact that after at least 40 years this issue is still not understood. The frame work for his paper is how to get exponential discs from numerical simulations of the hierarchical assembly of galaxies, a disc formation  mechanism very different to what Mestel envisaged and what is discussed here.

It is rather straight forward to see how the distribution of stars may be quite different to the distribution of the gas. For a Schmidt star formation law (Kennicutt, 2008) in which the Star Formation Rate per unit area (SFR) is proportional to some power of the gas surface density (SFR $\propto \Sigma_{gas}^{n}$ where $n \approx 1.4$) the stars and gas will have power law surface density slopes of $-n$ and -1 respectively. This simply assumes that the gas is distributed as in Mestel's disc and that star formation continues for the same length of time at all places in the disc. The real situation is obviously much more complex with the local stellar surface brightness dependent on the local physical conditions of the gas (propagation of a spiral density wave, for example) and the age of the stars. 

We conclude that the exponential nature of the surface brightness distribution of stars in disc galaxies provides no strong constraint on the overall distribution of the baryons because the star formation process is related to the gas surface density in a non-linear way.

\subsection{Measuring the local disc mass of our galaxy}
If the majority of the mass of galaxies is confined to the disc and not the halo then it should be apparent in derivations of the disc surface density. One way of measuring the local surface density of the Milky Way disc is to measure the scale height and velocity dispersion of stars. Bahcall (1984) did this and found a value of about 83 $M_{\odot}$ pc$^{2}$, Kuijken and Gilmore (1991) 71 $M_{\odot}$ pc$^{2}$, and Holmberg and Flynn (2004) 56 $M_{\odot}$ pc$^{2}$. With a surface density of about $10^{22}$ cm$^{-2}$ at 10 kpc from the galactic centre our `typical' model galaxy is well within these values at 76 $M_{\odot}$ pc$^{2}$. If we use a simple exponential model for the distribution of the stars above the plane then the predicted scale height is: $\beta=\frac{\sigma^{2}}{4\pi G \Sigma}$ where $\sigma$ is the velocity dispersion of the stars. With $\sigma=30$ km s$^{-1}$ and $\Sigma=10^{22}$ atoms cm$^{-2}$ the predicted scale height is a reasonable 0.2 kpc. 

van der Kruit and Searle (1982) have also found that the scale height of the stars is reasonably constant with radius. They argue that this is due to heating of the stars by interactions with giant molecular clouds and the passage of the spiral density wave. If true the stars are not actually particularly good test particles for measuring the surface mass density of the disc. But another possibility for the constant scale height is that the disc surface mass density, as in Mestel's disc, falls much more slowly with radius than the light from the stars would predict (Fig. 10). Olling (1995) has used the flaring of the HI disc at large galactic centre distances to show that the dark matter must reside in a flattened structure with axial ratio similar to a disc. In support of this Kalberla et al., (2007) say (they are modelling the atomic gas distribution of the Milky Way) that `..it is much easier to match a model to the observations if one considers dark matter associated with the Galactic disk.' 

\subsection{Disc stability}
That galaxies require massive halos, in addition to a disc, to give them stability has been another often quoted reason for invoking their existence. The origin of this idea can be traced back to a paper by Ostriker and Peebles (1973). They were carrying out a study of the stability of galactic discs using a numerical simulation. They found that models of the disc of our galaxy were `grossly' unstable to bar-like modes on very short timescales, but that models with, in their case a rigid, massive halo, were much more stable. They argued that a halo with a mass from 1-2 times that of the disc would provide the necessary stability, this is much smaller than the size of the halo now required to produce the observed rotation. Ostriker and Peebles (1973) suggested a stability criteria, based on the ratio of kinetic and potential energy of the disc and halo, that could be used to decide if a stable disc forms. Criteria like this have since been used by numerical modellers of galaxy populations so that they can identify in their codes dark matter halos that are firstly bound and then those that will form different types of galaxies - for example bar unstable galaxies are surprisingly often labelled as elliptical. It is not clear how useful these global criteria are in deciding whether a gas disc is actually stable or not within a simulation, or whether this is relevant to how a real galaxy evolves with time (Athanassoula, 2008).
For example Ostriker and Peebles (1973) suggested a massive halo as a means to suppress bar formation yet we now know that a large fraction of disc galaxies actually do have bars of one form or another. It is also quite clear from observations of the gas discs of galaxies that instability plays an important role in the way galaxies evolve and particularly how they form stars. Many disc galaxies are observed to have, along with bars, a large number of other distinct structures that form in the gas - spiral arms, filaments and clumps. If galaxies do have massive dark halos then they are not actually doing a very good job of preventing instabilities in the disc. 

Ostriker and Peebles (1973) actually showed that the discs they modelled became stable when the ratio of kinetic energy of rotation to gravitational energy was reduced to $t\approx0.14$. As this ratio for a 'typical' galaxy was higher than this they favoured a halo as a solution. For our `typical' galaxy $t \approx 0.5$, so at face value this disc is unstable because it is not hot enough. 

Efstathiou et al. (1982) questioned the validity of the Ostriker and Peebles stability criteria and suggested a new test:
\begin{center}
\large $t^{*}=v_{m} \left( \frac{R_{D}}{M_{D}G} \right)^{1/2} > 1.1$ \normalsize for stability against bar formation
\end{center}
where $v_{m}$ is the maximum velocity of the disc, $M_{D}$ is the disc mass and $R_{D}$ the exponential disc scale length. It is difficult to relate this to the Mestel disc as the criteria was developed for an exponential disc. However, $t^{*}$ looks like the square root of the kinetic energy per unit mass divided by the square root of the potential energy per unit mass in which case the value of $t^{*}$ would be about 1.4 for the our `typical' galaxy and the stability criteria would actually be satisfied. Subsequently Athanassoula (2008) has shown that this criterion ($t^{*}>1.1$) is also flawed and that it does not consistently in simulations, predict which galaxies will or will not produce strong bars. 

In summary it is not clear if any of these stability criteria tell us very much about whether a disc will maintain its gross properities over many rotation periods, which is our basic requirement. So, we do not believe that the stability of the disc provides a strong constraint on the necessity for an extensive massive halo (see also p. 603 Binney and Tremaine, 1994).

The Toomre instability (Toomre, 1964) is most often quoted as the means by which stars form, the criteria being:
\large
\begin{center}
$Q=\frac{c \kappa}{\pi G \Sigma} < 1$ 
\end{center}
\normalsize
where $\kappa$ is the epicyclic frequency and $c$ is the velocity dispersion of the gas (taken to be the sound speed $c=\sqrt{\frac{\gamma T}{m_{p}}}$). For a Mestel disc:
\large
\begin{center}
$Q=\left( \frac{4 \gamma kT}{\pi G m_{p} \Sigma_{0} R_{0}} \right)^{1/2}$ 
\end{center}
\normalsize
which interestingly is constant with radius and equal to 0.02 for our `typical' disc. So the prediction is that the disc will fragment and form stars, just as we would want it to do. Toomre stability requires a much hotter disc of about $10^{5}$ K.

If discs are to form in the way Mestel envisaged then initially there must have been an adiabatic collapse which gave rise to the gross features of the gas disc. This seems a reasonable assumption because in the early Universe we expect the gas to be of low metalicity and so there will be few metals to provide the necessary cooling lines. Subsequent radiative processes in metal enriched gas can produce the smaller scale structures and allow the gas to cool and form stars as the galaxy evolves. This is just the scenario proposed by those who first considered the collapse and subsequent fragmentation of proto-galactic gas clouds (Hoyle, 1953, Silk, 1977). The critical parameter (Silk, 1977) is the ratio of the cooling time ($t_{Cool} \propto \rho^{-1}$ where $\rho$ is the density) to the gravitaional collapse time ($t_{Grav} \propto \rho^{-1/2})$. For $t_{Cool}/t_{Grav} \propto \rho^{-1/2}>>1$ the collapse is adiabatic, but as the collapse continues, the density increases and the cooling time becomes shorter, eventually because of the nature of the hydrogen gas cooling curve the gas becomes isothermal and begins to fragment into clouds that eventually form stars.

\subsection{The gas consumption problem}
The average timescale for gas consumption by star formaton in our galaxy is of order 4 Gyr (Larson et al., 1980). From observations of the age, metalicity and velocities of local stars it appears that the star formation rate in the disc of our galaxy has been almost constant for billions of years longer than this. This almost constant star formation rate has recently been confirmed not just for our galaxy, but also for the Local Group as a whole by a detailed study of stellar populations (Drozdovsky et al., 2008). Surely as the gas is depleted the star formation rate should decrease? Larson et al.'s explanation, and that now normally accepted, is that there is continual infall of fresh gas into the galactic disc. The problem is that significant amounts of in falling gas with varying velocity and direction will thicken the galactic disc beyond what is observed (Quinn, 1987, Toth and Ostriker, 1992). This does not need to be the case if the stars already sit in a huge reservoir of gas in which they only represent a small fraction of the mass. 

\subsection{Constraints imposed by quasar absorption lines}
With typical atomic hydrogen column densities of $10^{22-23}$ cm$^{-2}$ (Fig. 10) sight lines to quasars through our model `typical' disc would produce damped Ly$_{\alpha}$ absorption spectra. Observations of damped Ly$_{\alpha}$ absorption systems produce column densities at most $10^{22}$ cm$^{-2}$. At first sight this seems a problem for our massive Mestel disc hypothesis. However, from the reddening of background quasars Fall et al. (1989) have shown that there is dust associated with Ly$_{\alpha}$ absorption features. Pettini et al., (1994) have shown that typical Ly$_{\alpha}$ absorption systems have metalicities about 0.1 of the solar value. If the dust is depleted in the same way and has the same properties as dust in the Milky Way then we might expect $A_{V} \approx \tau_{V}=\frac{N_{HI}}{10^{22}}$ where this is just a scaling of the Burstein and Heiles (1978) relation in which they measure approximately a V band optical depth of unity through an HI column density $(N_{HI})$ of 10$^{21}$ cm$^{-2}$. 
Thus optical sight lines to quasars through gas at column densities greater than about $10^{22}$ cm$^{-2}$ will be optically thick and these quasars will be missing from optically selected catalogues (Wright, 1990, Fall and Pei, 1993). Webster et al., (1995) have argued exactly this point. They show that radio selected quasars are red at optical wavelengths compared to optically selected quasars. They put this down to optical extinction by dust and infer that there is a large population of radio quiet quasars that remain undetected because they are extinguished. The high frequency of damped Ly$\alpha$ absorption features has normally been explained by assuming large extended HI halos at early epochs (Briggs and Wolfe 1983). This conclusion would need to be revisited if the actual column density through galactic discs is much higher than expected from the observed 21cm column densities of local galaxies.

\subsection{Dwarf galaxies}
A galaxy mass of $M \approx 10^{7}$ M$_{\odot}$ and a size of $R_{0} \approx 5$ kpc leads to a Mestel disc surface density of $\Sigma_{0} \approx 10^{19}$ cm$^{-2}$. Low mass galaxies like this must contain gas that is severely affected by the ionising background and no atomic hydrogen will be detected at 21cm. If the hydrogen is ionised then re-combination emission lines, particularly H$_{\alpha}$, should be observable in dwarf galaxies as a test of our hypothesis. This is a very similar experiment to that described by Bland-Hawthorn et al. (1997) in their search for ionised gas at the edge of nearby galaxies.

We know that at least back to a redshift of about $z=1$ the star formation rate per unit comoving volume was higher than it is today (Madau et al. 1998). As the Universe was also smaller the energy density of the ionising radiation must have been considerably higher. So, as the Universe ages from $z \approx 1$ the ionising flux decreases and the lower column densities of atomic hydrogen in low mass galaxies can resist ionisation. The HI can then cool and eventually form stars. If lower mass galaxies have lower column density discs ($\Sigma \propto M^{1/3}$) then there should be a relationship between their redshift and the ages of their stellar populations (downsizing?). We may also speculate that early on in the history of the Universe, before the majority of stars were formed, the ionising flux would again be low and all galaxies would be able to form stars, thus accounting for the old stellar population found in all dwarfs. 

It is observed that brighter dwarf galaxies have the lowest mass-to-light ratios something that is explainable by the baryons they contain (Mateo 1998). It is only for the lowest luminosity dwarf galaxies that the mass-to-light ratios become extreme, explained here by the onset of ionisation. Thus we suggest that there are three regimes: big (massive) galaxies appear to have high mass-to-light ratios because they have an as yet undetected cold gas component, intermediate galaxies have mass-to-light ratios fully accounted for by their baryons, small galaxies have high mass-to-light ratios because they are ionised. The boundaries being roughly at $10^{9}$ and $10^{7}$ M$_{\odot}$.

\section{The rotation curve puzzles and Tully-Fisher revisited}
With the optically thick Mestel disc illustrated in Fig 10 it is quite easy to see that the observed HI surface density is very close to a scaled version of the real surface density of hydrogen, thus explaining Bosma's (1978, 1981) observation that the rotation predicted from the gas is just a scaled version of the total. We are not suggesting that the gas is totally opaque and atomic, but that there could plausibly be very much higher surface densities of both atomic and molecular hydrogen than have previously been thought. 

For our hypothesised massive baryonic disc the largest fraction of the mass of a galaxy is now in the gas so there is no longer any reason to use maximum disc models based on the distribution of the stars. There is also now no longer any disc halo conspiracy because there is no longer any need for a halo. 

In the introduction we used a dynamical argument to try to explain the Tully-Fisher relation, this predicted a slope of 1/4. In fact the observed slope is closer to 1/3. Binney and Tremaine (1987) give exactly a value of 1/3 for the B band Tully-Fisher relation in precise agreement with what we found in section 3 (Fig. 7). Mestel shows that to form his disc via collapse of a spherical gas cloud, the density of the cloud ($\rho$) must be approximately constant. If this is true and about the same density for all galaxies then all Mestel discs will satisfy the Tully-Fisher relation $v_{m} \propto M^{1/3}_{Tot}$ (with luminosity as a good proxy for baryonic mass). Mestel discs are predicted to follow the Tully-Fisher relation with logarithmic slope of 1/3 because of the structure of the proto-galactic clouds they formed from. No conspiracy between surface brightness and mass-to-light ratio is required.

\section{Summary}
The observed rotation curves of galaxies do not necessarily imply that galaxies contain non-baryonic dark matter residing in an extended halo. The collapse of baryonic constant density/isothermal spheres leads naturally to surface density distributions that explain galactic rotation as long as one assumes that there is about a factor of 7 more baryonic material in the disc than is currently accounted for.
The observed almost constant surface density of atomic hydrogen may be explained by a combination of cold opaque atomic gas with additional cold molecular hydrogen and may be sufficient to account for this missing factor of 7. 

The massive baryonic disc has many attractive features. All `conspiracies' between the spherical dark matter halo and the baryonic disc are removed, it provides a reservoir of gas for continued star formation over a Hubble time (something possibly already identified in the $\gamma$-ray data) and it removes the necessity for unknown dark matter. However, currently there is no conclusive evidence for so many baryons in the disc and future observations of atomic and molecular hydrogen in galaxies are required to justify or refute our hypothesis. This though should be contrasted with the invention of dark matter and the continued lack of direct evidence for its existence. It should also be a concern that the additional baryonic mass required to account for rotation is just that required to reconcile observed baryons with those predicted by BBNS.

This paper does not try in any way to address other observations that imply the existence of dark matter, such as baryon acoustic oscillations, peaks in the cosmic microwave power spectrum and studies of large scale structure, though there are some who suggest that these too can be explained in other ways (Shanks, 2005). Here we only address the issue of the interpretation and implications of galaxy rotation curves. However we note in passing that other problems with the current $\Lambda$CDM model may have more straight forward solutions as a result:
\begin{enumerate}
\item Not enough low luminosity galaxies - if the major formation process is by monolithic collapse then this problem goes away as the mass spectrum of galaxies is a faithful reflection of the density fluctuations that created them. 
\item The cusp core problem - no dark matter halo so no problem.
\item Too many bright galaxies predicted by $\Lambda$CDM - see (i). above.
\item The `bias' problem is removed as baryons trace the mass.
\end{enumerate}
We conclude that it may be more productive to look for a dark baryonic component associated with hydrogen than to invoke and search for candidate exotic dark particles that have no other reason to exist other than to satisfy a whim or WIMP.

\begin{center}
{\bf References}
\end{center}
Ackermann M., 2011, ApJ, 726, 81 \\
Ade et al., 2011, arXiv1101.2029 \\
Allen R., 2002, in 'Seeing Through the Dust: The Detection of HI and the Exploration of the ISM in Galaxies', ASP Conf. Proc., Vol. 276, Edited by A. R. Taylor, T. L. Landecker, and A. G. Willis, Pub. ASP, p.288
Athanassoula E., 2008, MNRAS, 390, 69\\
Babcock H, 1939, Lick obs bulletin No 498, p. 41 \\
Bahcall J., 1984, ApJ, 287, 926\\
Benson A., Lacey C., Frenk C., Baugh C. and Cole S., 2004, MNRAS, 351,1215 \\
Bigiel et al., 2010, AJ, 140, 1194 \\
Binney J. and Tremain S., 1994, In 'Galactic Dynamics', pub. Princeton University Press \\
Bland-Hawthorn J., Freeman K. and Quinn P., 1997, ApJ, 490, 143\\
Bohlin R., Savage B. and Drake J., 1978, ApJ, 216, 291\\
Boselli A., Lequeux J. and Gavazzi G., 2002, A\&A, 384, 33 \\
Bosma A., 1978, PhD Thesis, University of Groningen\\
Bosma A., 1981, AJ, 86, 1791\\
Broeils A. and Rhee M., 1997, A\&A, 324, 877 \\
Briggs F. and Wolfe A., 1983, ApJ, 268, 76\\
Braun R., Thilker D., Walterbos R. and Corbelli E., 2009, ApJ, 695, 937 \\
Burgh E., France K., and McCandliss S., ApJ, 658, 446 \\
Burstein D. and Heiles, C., 1978, ApJ, 225, 40\\
Carignan C., Charbonneau P., Boulanger F. and Viallefond F., 1990, A\&A, 234, 43 \\
Crampin D. and Hoyle F., 1964, ApJ, 74, 186\\ 
de Blok W. et al., 2008, AJ, 136, 2648\\
de Blok W., Bosma A. and McGaugh S., 2003, MNRAS, 340, 657 \\
Denham W. and Binney J., 1998, MNRAS, 294, 429\\
Dickman R., 1978, ApJSS, 37, 407 \\ 
Dickey J., Salpeter E. and Terzian Y., 1978, ApJSS, 36, 77\\
Dickey J. et al., 2009, ApJ, 693, 1250 \\
Doroshkevich A., Tucker D., Allam S. and Way M., 2004, AA, 418, 7\\
Draine B. et al., 2007, ApJ, 663, 866 \\
Dutton A., 2009, MNRAS, 396, 121 \\
Efstathiou G., Lake G. and Negroponte J., 1982, MNRAS, 199, 1069\\
Eggen O., Lynden-Bell D. and Sandage  A., 1962, ApJ, 136, 748\\
Fall S. and Pei Y., 1993, ApJ, 402, 479\\
Fall S., Pei Y. and McMahon, 1989, ApJ, 341, 5\\
Freeman K., 1970, ApJ, 60, 811\\
Freeman K., 1992, In 'Physics of Nearby Galaxies, Nature or Nurture', Ed. T. Thuan and J. Balkowski, Editions Frontiers, Gif-sur-Yvette, p. 201 \\
Fuchs B., 2002, In 'Dark matter in astro- and particle physics. Proceedings of the International Conference DARK 2002, Cape Town, South Africa, 4 - 9 February 2002. Ed. H. V. Klapdor-Kleingrothaus and R. D. Viollie, Pub. Springer, p. 28  \\
In 'Dark matter in astro- and particle physics. Proceedings of the International Conference DARK 2002, Cape Town, South Africa, 4 - 9 February 2002. H. V. Klapdor-Kleingrothaus, R. D. Viollier (eds.). Physics and astronomy online library. Berlin: Springer, ISBN 3-540-44257-X, 2002, p. 28 - 35 \\
Fukugita M., Hogan C. and Peebles P., 1998, ApJ, 503, 518 \\
Garwood R. and Dickey J., 1989, ApJ, 338, 841 \\
Gibson S., Taylor A. and Higgs L., 2005, ApJ, 626, 195\\
Heiles C. and Troland T., 2003, ApJS, 145, 329\\
Hessmann F. and Ziebart M., A\&A, 523, 121 \\
Holmberg J. and Flynn C., 2004, MNRAS, 352, 440\\
Hoyle F., 1953, ApJ, 118, 513\\
Hunter J., Ball R. and Gottesman S., 1984, MNRAS, 208,1\\
Israel, F., 2000, In 'Molecular hydrogen in space',  Ed. F. Combes, and G. Pineau des Forts, Pub. Cambridge University Press, p. 326 \\
Kalberla P., Dedes L., Kerp J. and Haud U., 2007, AA, 469, 511\\
Kalberla P. and Kerp J., 1998, A\&A, 339, 745 \\
Kalberla P., 2003, ApJ, 588, 805 \\
Kalberla P., 2004, Ap\&SS, 289, 239 \\
Kalnajs A., 1987,in 'Dark Matter in the Universe, IAU Sym. 117, Ed. J. Kormendy and G. Knapp, Reidel, Dordrecht, P. 28\\
Kennicutt R., 1989, ApJ, 344, 685 \\
Kennicutt R., 2008, In 'Pathways Through an Eclectic Universe', ASP Conference Series, Vol. 390, Ed.J. H. Knapen, T. J. Mahoney, and A. Vazdekis, p. 149
Kerr F., 1975, In 'Stars and Stelar Systems', Vol IX, Ed. A. Sandage\\
Knee L. and BruntC., 2001, Nature, 412, 308 \\
Kolpak, M., Jackson J., Bania T. and Dickey J., 2002, ApJ, 578, 868 \\
Kopp M., Roueff D and Pineau des Forets G., 2000, MNRAS, 315, 37 \\
Kuijken K. and Gilmore G., 1991, ApJ, 367, L9\\
Larson R., Tinsley B. and Caldwell C., 1980, ApJ, 237, 692\\
Lin D. and Pringle J., 1987, ApJ, 320, L87 \\
Lequeux J., Allen R. and Guilloteau S., 1993, A\&A, 280, 23 \\
Madau P., Pozzetti L. and Dickinson M., 1998, ApJ, 498, 106 \\
Magrini L. et al., 2001, A\&A, in press \\
Maloney P., 1993, ApJ, 414, 41\\
Mestel L., 1963, MNRAS, 126, 553\\
Minchin et al., 2004, MNRAS, 355, 1303\\
Navarro J. Frenk C. and White S., 1994, MNRAS, 267, L1\\Pfenniger
Navarro J., Frenk, C. and White S., 1996, ApJ, 463, 563 \\
Olling R., 1995, AJ, 110, 591\\
Ostriker J. and Peebles P., 1973, ApJ, 186, 467\\
Pagel B., 1999, 'The Low Surface Brightness Universe", ASP Conference Series 170, Ed. J.I. Davies, C. Impey, and S. Phillipps, Astr, p.375 \\
Palunas P. and Williams T., AJ, 120, 2884\\
Pettini M., Smith L., Hunstead R. and King D., 1994, ApJ, 108, 2046\\
Pfenniger D., Combes F. and Martinet L., 1994, AA, 285, 79\\
Pfenniger D. and Combes F., 1994, AA, 285, 94\\
Pildis R. and McGaugh S., 1996, ApJ, 470. L77 \\
Portas A. et al., 2009, In 'The Galaxy Disk in Cosmological Context', Proceedings of IAU Symposium 254. Ed. J. Andersen, J. Bland-Hawthorn and B. Nordstršm. Pub. Cambridge University Press, p.52 \\
Quinn P., 1987, In 'Nearly Normal Galaxies', Ed. S. Faber, Springer, New-York, p. 138\\
Rees M. and Ostriker J., 1977, MNRAS, 179, 541\\
Revaz Y., Pfenniger D., Combes F. and Bournaud F., A\&A, 501, 171 \\
Richter P., Savage B., Sembach K. and Tripp T., 2006, A\&A, 445, 827 \\
Richter P., Savage B., Tripp T. and Sembach K., 2004, ApJSS, 153, 165 \\
Roberts M. and Rots A., 1973, AA, 26, 483\\
Rogstad D. and Shostak G., 1972, AJ, 176, 315\\
Rubin C., Ford, W. K., Jr. and Thonnard, N., 1978, 225, L107\\
Seiden P., Schulman L. and Elmegreen B., 1984, ApJ, 282, 95 \\
Shanks T., 2005, In 'Maps of the Cosmos', Proc. of IAU Sym 216,. Ed. M. Colless, L. Staveley-Smith and R. Stathakis. Pub Astronomical Society of the Pacific, p. 398\\
Silk J., 1977, ApJ, 211, 638 \\
Solomon P. and Barrett J., 2001, In 'Dynamics of Galaxies and their Molecular Cloud Distributions', Ed. F. Combes and F. Casoli, IAUS, 146, 235 \\
Spergel et al., 2007, ApJS, 170, 377\\
Stark et al., 1992, ApJS, 79, 77\\
Steinmetz M. and Navarro J., 1999, ApJ, 513, 555\\
Strong A., Moskalenko I., Reimer O., Digel S. and Diehl R., 2004, A\&A, 422, L47 \\
Takei Y., 2011, arXiv:1011.2116\\
Toomre A., 1964, ApJ, 139, 1217 \\
Toth G. and Ostriker J., 1992, ApJ, 389, 5\\
Tully R. and Fisher J., AA, 54, 661\\
van Albada T. and Sancisi R., 1986, Royal Society Discussion on 'The Material Content of the Universe', Philosophical Transactions, Series A, vol. 320, no. 1556, p. 447\\
Valentijn E. and van der Werf P., ApJ, 522, L29 \\
van der Kruit P and Searle, 1982, AA, 110, 61\\
van Dishoeck E. and Black J., 1988, ApJ, 334, 771 \\
Walter et al., 2008, AJ, 136, 2563\\
Webster, R., Francis, P., Peterson, B., Drinkwater, M. and Masci, F., 1995, Nature, 375, 469 \\
Wilson C., 1995, ApJ, 448, L97 \\
White M. and Rees M., 1978, MNRAS, 183, 341\\
Wright, E., 1990, ApJ, 353, 411\\
Yoshii Y. and Sommer-Larson J., 1989, MNRAS, 236, 779 \\
Zwaan M., van der Hulst J., de Blok E. and McGaugh S., 1995, MNRAS, 273, L35\\

\end{document}